\documentclass[twocolumn]{aastex62}

\usepackage{verbatim}
\usepackage[usenames, dvipsnames]{}

\newcommand{\gasyield}{\ensuremath{105\pm8}}
\newcommand{\wateryield}{\ensuremath{5\pm2}}
\newcommand{\kepleryield}{\ensuremath{20\pm3}}
\newcommand{\lateyield}{\ensuremath{37\pm6}}

\shorttitle{Exoplanet Science for EVE}
\shortauthors{EVE Exoplanet Science Team}

\begin{document}

\title{Preparing for the Early eVolution Explorer: Detecting the Primordial, Transiting Exoplanet Population }

\author[0000-0002-4891-3517]{George Zhou}
\affiliation{University of Southern Queensland, Centre for Astrophysics, West Street, Toowoomba, QLD 4350 Australia}

\author[0000-0001-7615-6798]{James G. Rogers}
\affiliation{Institute of Astronomy, University of Cambridge, Madingley Road, Cambridge CB3 0HA, United Kingdom}

\author[0000-0002-0040-6815]{Jennifer~A.~Burt}
\affiliation{Jet Propulsion Laboratory, California Institute of Technology, 4800 Oak Grove Drive, Pasadena, CA 91109, USA}

\author[0000-0002-1228-9820]{Eve J. Lee}
\affiliation{Department of Astronomy \& Astrophysics, University of California, San Diego, La Jolla, CA 92093-0424, USA}

\author[0000-0001-9158-9276]{Sydney Vach}
\affiliation{University of Southern Queensland, Centre for Astrophysics, West Street, Toowoomba, QLD 4350 Australia}
\affiliation{European Southern Observatory, Karl-Schwarzschild-Str. 2, 85748 Garching bei München, Germany}

\author[0000-0002-3656-6706]{Ann Marie Cody}
\affiliation{SETI Institute, 339 N Bernardo Ave, Suite 200, Mountain View, CA 94043, USA}

\author{Mark Swain}
\affiliation{Jet Propulsion Laboratory, California Institute of Technology, 4800 Oak Grove Drive, Pasadena, CA 91109, USA}

\author[0000-0001-8292-1943]{Neal J. Turner}
\affiliation{Jet Propulsion Laboratory, California Institute of Technology, 4800 Oak Grove Drive, Pasadena, CA 91109, USA}

\author[0000-0003-3654-1602]{Andrew W. Mann}
\affiliation{Department of Physics and Astronomy, The University of North Carolina at Chapel Hill, Chapel Hill, NC 27599, USA}

\author[0000-0002-8399-472X]{Madyson G. Barber}
\affiliation{Department of Physics and Astronomy, The University of North Carolina at Chapel Hill, Chapel Hill, NC 27599, USA}

\author[0000-0002-5258-6846]{Eric Gaidos}
\affiliation{Department of Earth Sciences, University of Hawai'i at Mānoa, Honolulu, HI 96822, USA}
\affiliation{Institute for Astrophysics, University of Vienna, A-1180 Wien, Austria}

\author[0000-0002-0583-0949]{Ward Howard}
\affiliation{Department of Astrophysical and Planetary Sciences, University of Colorado, 2000 Colorado Avenue, Boulder, CO 80309, USA}

\author[0000-0002-4115-0318]{Laura Venuti}
\affiliation{SETI Institute, 339 N Bernardo Ave, Suite 200, Mountain View, CA 94043, USA}
\affiliation{Visiting Fellow, School of Physics, UNSW Science, Kensington, NSW 2052, Australia}

\author{Damon F. Landau}
\affiliation{Jet Propulsion Laboratory, California Institute of Technology, 4800 Oak Grove Drive, Pasadena, CA 91109, USA}

\author{Valerie Scott}
\affiliation{Jet Propulsion Laboratory, California Institute of Technology, 4800 Oak Grove Drive, Pasadena, CA 91109, USA}

\author{Alan Didion}
\affiliation{Jet Propulsion Laboratory, California Institute of Technology, 4800 Oak Grove Drive, Pasadena, CA 91109, USA}

\author{David Makowski}
\affiliation{Jet Propulsion Laboratory, California Institute of Technology, 4800 Oak Grove Drive, Pasadena, CA 91109, USA}

\author{Jamie Nastal}
\affiliation{Jet Propulsion Laboratory, California Institute of Technology, 4800 Oak Grove Drive, Pasadena, CA 91109, USA}

\author{Evgenya L. Shkolnik}
\affiliation{School of Earth and Space Exploration, Arizona State University, Tempe, AZ 85281, USA}

\author{Meredith A. MacGregor}
\affiliation{Department of Physics and Astronomy, Johns Hopkins University, 3400 N Charles St, Baltimore, MD 21218, USA}

\begin{abstract}
The close-in small planet population may be formed either with hydrogen/helium dominated envelopes or with water-rich interiors. Both scenarios reproduce the present day planet population in mass, radius, and periods, and are difficult to differentiate with the mature planet demographic. Hydrogen/Helium `gas-dwarfs' have low mean molecular weight atmospheres, while `water-worlds' have envelopes that are significantly heavier, and as such these two scenarios have different evolution tracks that diverge in the first $\approx 50$\,Myr of their evolution. We show that a  low Earth orbit multi-band photometric survey mission, within the scope of the NASA Small Explorers Program (SMEX), can determine the frequency of young close-in planets at the 5\% level and definitively differentiate between the competing `gas-dwarf' and `water-world' hypotheses. We simulate a 2.5 year mission capable of simultaneous multi-band near-ultraviolet (NUV), optical, and near infrared (NIR) wide field photometry. Such a mission would perform a photometric survey of $30$ different stare-fields selected to probe the young star population. The mission will yield $\approx 100$ transiting planets in young star clusters and associations with ages $<50$ Myr. In comparison, only 20 such planets are known from \emph{K2} and \emph{TESS} today. 
\end{abstract}


\section{Introduction \label{sec:intro}}

\textit{Kepler} revealed that the distribution of planets with sizes $\lessapprox 4$~R$_\oplus$ and orbital periods $\lessapprox 100$~days is bimodal, divided by a deficit of planets centered at $\approx 1.8 R_\oplus$, known as the radius valley \citep[e.g.][]{Fulton2017,Petigura2022}. Planets below the valley, referred to as `super-Earths', have bulk densities (from photometric transit-derived radii and Doppler radial velocity-based masses) consistent with an Earth-like composition, while planets above the valley, referred to as `sub-Neptunes', have reduced bulk densities that suggest they host substantial atmospheres \citep[e.g.][]{Wu2013,Rogers2015,Wolfgang2016,Luque2022}. 

The dichotomy in composition between planets above and below the radius valley has been explained by two hypotheses that predict planet radial evolution at differing timescales. The `gas-dwarf' scenario posits that both the super-Earth and sub-Neptune populations formed through a unified formation channel, in which rocky cores accreted hydrogen/helium-rich envelopes from their natal protoplanetary disks \citep[e.g.][]{Lee2014,Ginzburg2016}. Then, within the first hundred million years, atmospheric escape, such as boil-off, core-powered mass loss and XUV photoevaporation\footnote{as per \citet{Owen2012}, from $\approx1-2$keV in X-ray to $>13.6$eV in the extreme-UV.}, strip the primordial envelopes from a subset of planets with lower core masses and that experience higher irradiation \cite[e.g.][]{Owen2013,Lopez2013,Owen2016b,Ginzburg2016,Ginzburg2018,Gupta2019,Rogers:2021,Rogers2024a,2025ApJ...985..100A}. The removal of a hydrogen/helium envelope significantly reduces the observed radius of a given planet. As a result, the radius valley is carved by atmospheric escape, with the population of stripped planets forming the super-Earths below the valley. 

Within the gas-dwarf hypothesis also exists the possibility that planet cores may only grow to be of sufficient mass for gas accretion only after most of the gas disk has been dissipated, and orbit crossings become common. The radius valley may form naturally as a result of a threshold for gas accretion; only cores of a sufficient mass can trigger accretion and become sub-Neptune in size \citep{2016ApJ...817...90L,2021ApJ...908...32L,2022ApJ...941..186L}. The planets may then be subjected to additional photoevaporative evolution, further carving the radius valley for the nearest planets. 

The second hypothesis to explain the radius valley is the `water-world' scenario. In this model, super-Earths and sub-Neptunes follow differing formation and evolution pathways. As a result, they have fundamentally different internal compositions and sizes, which gives rise to the radius valley \citep{Luque2022}. 
Under this hypothesis, super-Earths have Earth-like densities and composition throughout, while sub-Neptunes have a dense rocky interior, and an envelope of water that make up tens of per-cent in total planet mass. \citep[e.g.][]{Mordasini2013,Venturini2017,Zeng2019,Burn2024}.

\begin{figure}
    \centering
    \includegraphics[width=1\linewidth]{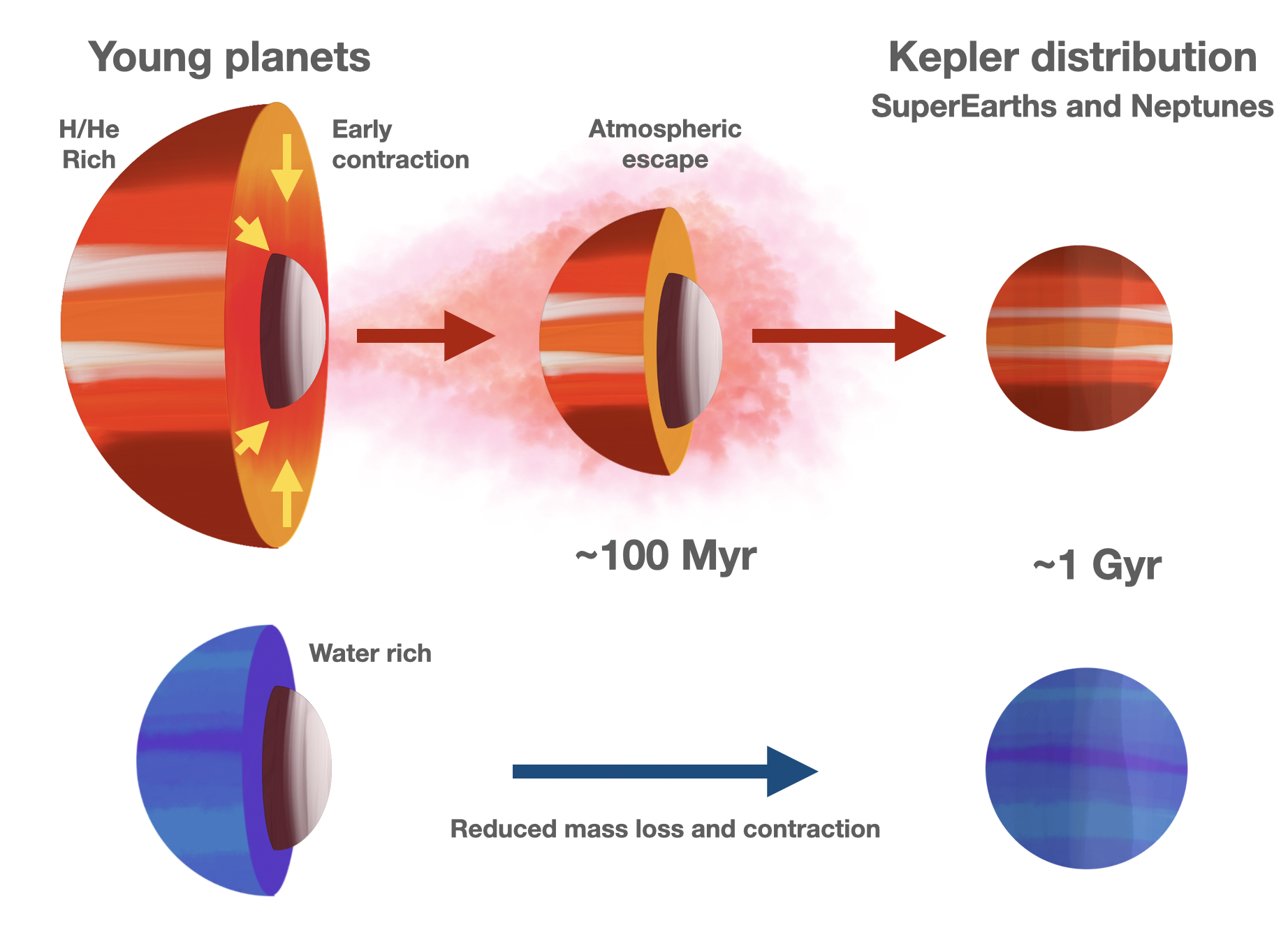}
    \caption{The small planet population can be explained by competing hypotheses. Planets may form with substantial, low mean-molecular weight atmospheres (gas-dwarfs). Depending on when the gas envelope was accreted, these planets may undergo rapid radial evolution during the first hundred million years due to the combined atmospheric escape effects of boil-off, contraction, and photo-evaporation. Planets may also form with volatile rich, higher mean molecular weight envelopes (water-worlds). Such planets undergo far less radial evolution within this time frame. The densities and radii of two populations are degenerate by the mature ages sampled by the \emph{Kepler} and \emph{TESS} missions. Targeting planets around known star forming regions and young associations can statistically differentiate between these two scenarios. }
    \label{fig:illustration}
\end{figure}

Our ability to distinguish between these formation scenarios is hindered by a well-established degeneracy in determining a planet's interior composition given only its bulk density (from mass and radius) \citep[e.g.][]{Rogers2015,Bean2021,Rogers:2023,Rogers2025a}. As such, `gas-dwarfs' and `water-worlds' can both explain the majority of the super-Earth and sub-Neptune populations in mass, radius, and period.

A key distinction between the models, however, is the order-of-magnitude difference in predicted envelope mean molecular weight. Gas-dwarfs initially accrete low mean molecular weight hydrogen/helium dominated envelopes from their nascent protoplanetary disks, which may be increased by interior-atmosphere chemical reactions to within an approximate range of $\approx 2.3 - 5$ atomic mass units (amu) \citep[e.g.][]{Schlichting2022}. Water-worlds, on the other hand, host water-rich, high mean molecular weight envelopes in the approximate range of $\approx 10 - 18$ amu \citep[e.g.][]{Burn2024,Burn2024b}.

Numerous direct observations of escaping hydrogen and helium from sub-Neptunes are evidence in favor of the gas-dwarf scenario \citep[e.g.][]{2018A&A...620A.147B,2015Natur.522..459E,2016A&A...591A.121B,2018ApJ...868L..34M,2020A&A...634L...4D,2025Natur.638..636L,2025ApJ...985L..10A}. However, JWST atmospheric observations have so far revealed a mixed picture, with some sub-Neptunes interpreted to host high mean molecular weight atmospheres, e.g., TOI-836 c, \citep{2024AJ....168...77W}; GJ 9827 d, \citep{2024arXiv241003527P}; and GJ 1214 b, \citep{2023ApJ...951...96G}, while others are interpreted to host low mean molecular weight atmospheres, e.g., TOI-421 b \citep{Davenport2025}, and TOI-270 d \citep{Benneke2024}. Of note, some young planets have been characterized to have low bulk densities with low mean molecular weight atmospheres, providing perhaps supporting evidence for the existence of `gas-dwarfs', e.g., the V1298 Tau system \citep{Barat:2024a,Barat:2024b} and HIP 67522 b \citep{Thao:2024}. 

\citet{Rogers2025b} showed that a promising strategy to determine which of the gas-dwarf or water-world formation scenarios is dominant is by measuring the radius of a population of transiting exoplanets with ages $\lesssim 100$~Myrs. While each scenario can reproduce the observed population of sub-Neptunes around mature stars ($\gtrsim 1$~Gyr), the sizes of sub-Neptunes shortly after protoplanetary disk dispersal are strongly controlled by the bulk mean molecular weight of their envelopes. Since `gas-dwarfs' host low mean molecular weight envelopes, their radii should be significantly larger than `water-worlds' of equivalent age and mass. The radii of `water-worlds' do not undergo significant contraction throughout their lifetime (Figure~\ref{fig:illustration}). This difference in planet radius evolution allows one to differentiate between models of formation that result in planets today with low and high mean molecular weight envelopes.

Constructing a statistically useful catalog of young planets with existing facilities is challenging due to the larger pre-MS radii and elevated magnetic activity, thereby increased photometric variability of the young host stars. Only 20 known transiting planets have ages $<50$ Myr\footnote{IRAS 04125+2902 b \citep{2024Natur.635..574B}, K2-33 \citep{David2016,Mann:2016}, TOI-1227 \citep{2022AJ....163..156M}, TIC 88785435 \citep{2025AJ....170..131V}, HIP 67522 \citep{Rizzuto2020,2024ApJ...973L..30B}, AU Mic \citep{Plavchan2020}, V1298 Tau \citep{David2019b,David2019}, HD 109833 \citep{2023AJ....165...85W}, NGTS-33 \citep{2025MNRAS.536.1538A}, TOI-6448 \citep{2026AJ....171...20B}, TOI-837 \citep{2020AJ....160..239B}, DS Tuc Ab \citep{2019ApJ...880L..17N}}. Statistical study of these known planets are insufficient to definitively break the degeneracy of the planet formation models \citep{2024AJ....167..210V}.   
A majority of these planets were identified via observations of moving groups within 150 pc from \emph{K2} and \emph{TESS}, including the 1--15\,Myr old Taurus-Auriga star-forming region at $\sim 150$\,pc \citep{David2019,David2019b,2024Natur.635..574B}, the 10-20\,Myr old Scorpius-Centaurus complex at 130\,pc \cite[e.g.][]{Mann:2016,David:2016,Rizzuto:2020,2022AJ....163..156M,2022A&A...667L..14Z,2025AJ....170..131V}, as well as more sparse regions yielding singular transiting planet systems, such as the AU Mic system in the $\beta$ Pictoris moving group at 30-50\,pc \citep{Plavchan2020}. The majority of stars in any star-forming region is naturally biased towards M-dwarfs due to the initial mass function. To find the most number of planets in such regions requires sensitivity to late-type stars. As such the effectiveness of \emph{TESS} to detect a statistically significant population of young planets is limited by the number of young associations within $\sim 150$\,pc.

This paper examines the capabilities of a wide field photometric mission dedicated to surveying star forming regions and young stellar associations, fitting within the scope of NASA SMEX. We perform simulations to demonstrate that a conceptual SMEX-class mission could detect enough transiting planets $<50$Myr old to discriminate between the gas-dwarf and water-world formation scenarios. One key to success is the simultaneous multi-band photometric monitoring -- the trio of NUV, optical, and NIR channels. Photometric transit detections will be performed in the optical and NIR bands. The NIR band captures nearly half the flux from cool newly formed M-stars, and also provides power in differentiating between astrophysical false-positive scenarios without the need for additional ground-based photometric follow-up. The NUV band will provide simultaneous flare detection and modeling, helping to reduce the influence of activity as one noise source for the detection of planets around young stars.

\section{Survey strategy and simulations \label{sec:simulations}}

\subsection{The Early eVolution Explorer \label{sec:eve}}
This paper is inspired by the design of the Early eVolution Explorer mission concept \citep[EVE;][]{2025AAS...24523409M}, a NASA SMEX class mission concept that aims to examine the physical processes that shape the inner disk architectures, quantify the photochemical impacts of stellar flares on young planet atmospheres, and detect a primordial transiting planet population about newly formed stars. 

Within the scope of a SMEX mission, we perform simulations for a mission design that will have a $5^\circ\times5^\circ$ field of view (FoV) and three photometric bandpasses; a near UV channel from 200 -- 300\,nm with a 12.5\,cm aperture, an optical channel from 500 -- 900\,nm with a 20\,cm aperture, and a NIR channel has an aperture of 18\,cm and covers the 1.0-1.6\,$\mu$m wavelength range. The optical and NIR channels have point response functions with angular width of at most 15\arcsec. These bands are chosen such that the telescopes and detectors can achieve similar fields of view and spatial scale within one SMEX-class spacecraft. Exoplanet transit detections will be made by combining the optical and NIR channels, while the NUV channel will be used to detect, model, and mitigate the effects of stellar flares on the detection of planetary transits as per \citet{2025AJ....169...27H}.  

EVE will follow a polar orbit just over 1,200\,km in altitude that minimizes geocoronal interference and radiation doses while maximizing access to the sky and solar exposure. The orbit provides sustained access to two 25-degree radius fields of regard centered at 0 degrees declination and 103 and 283 degrees in right ascension (Figure~\ref{fig:maps}). 

Over the course of its 3 year primary mission, EVE will survey thirty $5^\circ\times5^\circ$ fields in young stellar clusters that together contain $\approx 20,000$ stars with ages $\lesssim 50$\,Myr. These star-forming regions are too distant for existing missions such as \emph{TESS} to detect small planets in, but can yield a significant planet sample that will unveil the primordial demographics of the small planet population. Each field will be continuous and simultaneously monitored in NUV/Optical/NIR pass-bands over 30\,days 

\subsection{Survey target fields \label{sec:targetfields}}

We compiled an ensemble list of 1.1 million kinematically associated stars from literature. This target list is collated from catalogs published in \citet{Kounkel2018}, \citet{Kos2018}, \citet{CantatGaudin2019}, \citet{Tang2019}, \citet{Lodieu2019}, \citet{McBride2019}, \citet{Roeser2019}, \citet{CantatGaudin2020}, \citet{Chen2020}, \citet{Kounkel2020}, \citet{Grasser2021}, \citet{Jerabkova2021}, \citet{Krolikowski2021}, \citet{Kos2021}, \citet{Pavlidou2021}, \citet{Esplin2022}, \citet{Luhman2022}, \citet{Newton2022}, \citet{Nunez2022}, \citet{Pang2022}, \citet{Luhman2023}, \citet{Olivares2023}, \citet{Ratzenbock2023}, \citet{SanchezSanjuan2024}, \citet{Zerjal2024}, and \citet{2025A&A...702A..63R}. We cross matched this list against the \emph{TESS} Input Catalog (TIC), from which we adopted stellar parameters \citep{Stassun2019} for the masses and radii of our target list for the remainder of this analysis\footnote{In-depth spectral energy distribution analyses of planet-hosting pre-main sequence stars have generally shown consistency between the TIC and published results in effective temperature, mass, radii estimates \citep[e.g.][]{Mann:2016, Rizzuto:2020,Plavchan2020,2022AJ....163..156M}}. 

\begin{figure*}
    \centering
    \includegraphics[width=0.9\linewidth]{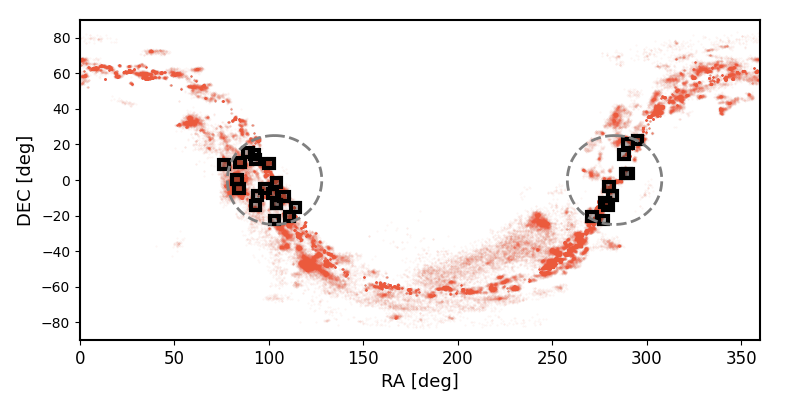} 
    \caption{Survey fields of EVE aligned to the distribution of young stars. EVE will survey $\approx 30$ separate pointings over a 2.5 year primary mission. Stars with ages $<50\,\mathrm{Myr}$ are plotted in the background. The 25\,degree circles mark the fields of regard for the mission, within which sustained observations are possible.}
    \label{fig:maps}
\end{figure*}

Initial estimates of planet yield (described in Section~\ref{sec:transit_detection}) were conducted to determine which $5^\circ\times5^\circ$ regions of the sky yielded the highest number of detectable young planets. These fields are then manually vetted to confirm they can be sampled over the course of a 2.5 year mission with stare durations of $30\,$days per field. An example set of pointings for such a mission concept is shown in Figure~\ref{fig:maps}. This example set of pointings encompass a primary mission survey of $\sim20,0000$ stars with ages $\lesssim 50$\,Myr that can be monitored over thirty individual $5\times5^\circ$ fields. We note that the Orion star forming complex was identified as the most important for such a mission's exoplanet yield, given its high stellar density, and that it is too distant $(\sim 400\,\mathrm{pc})$ to yield a significant number of planets from the small aperture of \emph{TESS}. As such, the Orion complex will likely deserve additional efforts to maximize its coverage in a mission design. We note that numerous additional factors, such as final orbit selection, and eventual launch date, will affect the true set of fields a Low Earth Orbit mission can survey. 

\begin{figure*}
    \centering
    \includegraphics[width=1\linewidth]{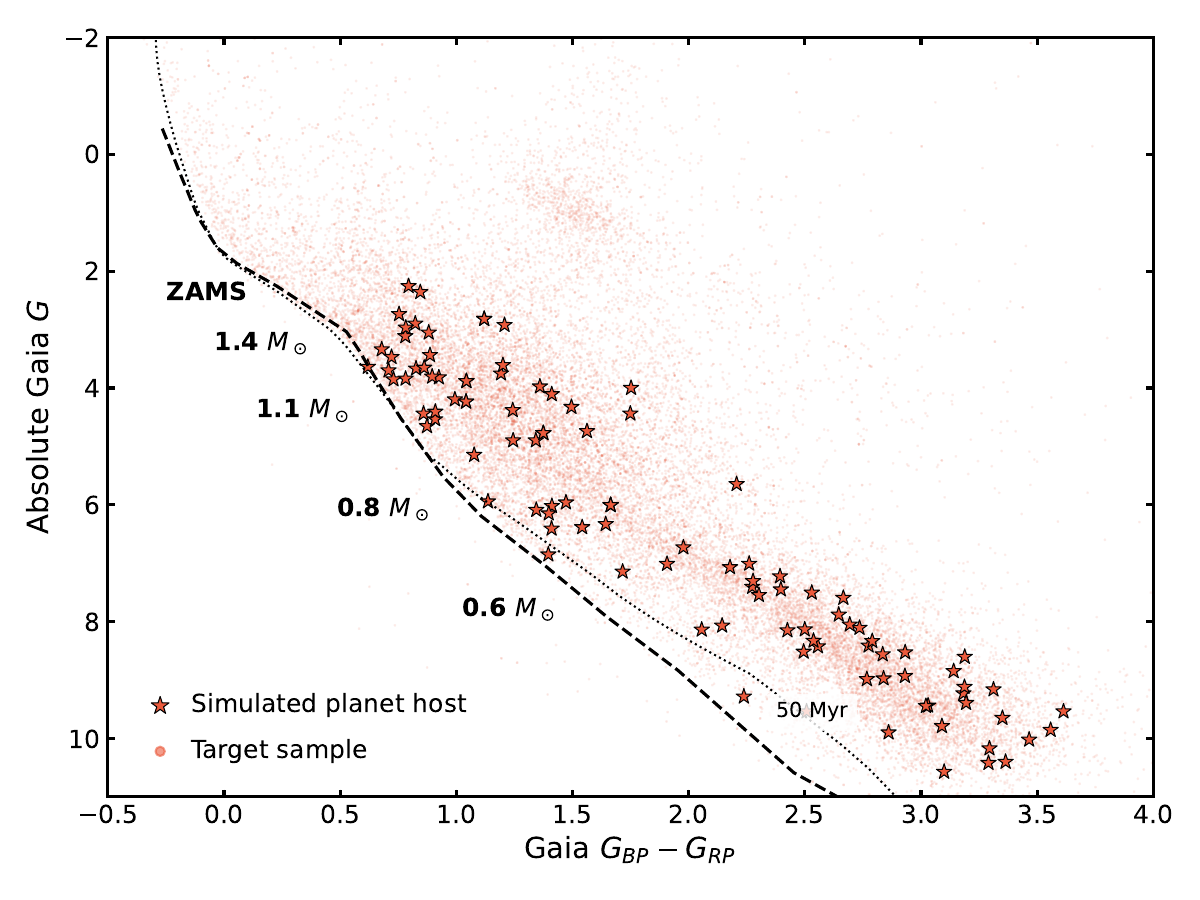}
    \includegraphics[width=1\linewidth]{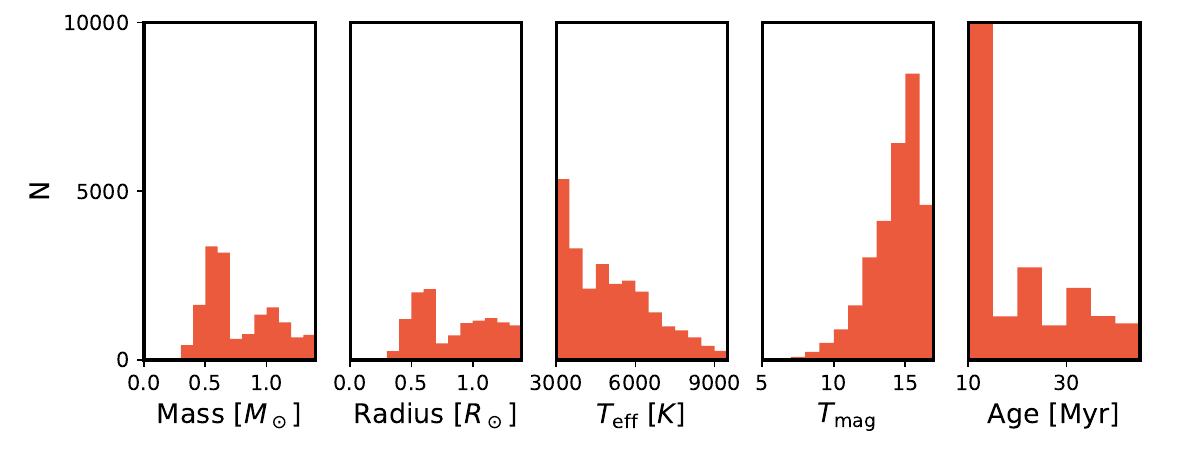}
    \caption{The stellar property distribution encompassed within the EVE fields plotted as over the color-magnitude diagram (top), and as a function of stellar mass, radius, effective temperature, magnitude, and age (bottom). The zero-age main sequence (ZAMS) and the 50\,Myr solar metallicity non-rotating isochrone as per \citet{Dotter2016} are marked by the dashed and dotted lines respectively. The simulated planet hosts are marked by stars. Stellar properties are adopted from the TESS input catalog \citep{Stassun2019}.}
    \label{fig:stellarpop}
\end{figure*}

The properties of stars surveyed over the 2.5 year primary mission are shown in Figure~\ref{fig:stellarpop}. The target list is dominated by late-type stars, as is dictated by the initial mass function and the pre-main sequence evolution tracks. Section~\ref{sec:transit_detection} shows that the combined NUV, optical, and NIR observations will have the capability to detect planets around pre-main-sequence, late-type stars as faint as a \emph{TESS} magnitude of $T_\mathrm{mag} \approx 16$\footnote{95\% of planet discoveries will be around stars brighter than $T_\mathrm{mag} = 16$}. 

We note that many stars in star forming regions are irregular variables, including dippers, bursters, amongst others. These irregular light curves likely prohibit the detection of transiting planets. One estimate for the frequency of such stars is the fraction of stars hosting disks. Estimates of disk fraction within the Orion star forming region range from 10-30\% \citep[e.g.][]{2007ApJ...662.1067H,2007ApJ...671.1784H,2012AJ....144..192M}, and range between 5-10\% for the Scorpius Centaurus complex \citep[e.g.][]{2016MNRAS.461..794P}. \citet{2018AJ....156...71C} notes most disk-bearing stars exhibit significant stellar variability in \emph{K2} light curves of the Upper Sco--Oph region. Of which, $\approx 30$\% of disk-bearing stars exhibit dipper-like light curve behavior, and an additional $\approx 10$\% exhibit outbursts that will also prevent detections of transiting planets. Similarly, \emph{CoRoT} observations of the star forming region NGC 2264 showed $\approx 30\%$ of disk-bearing stars to exhibit dipper behavior. These stars have not been excluded from our planet-yield, but will likely impact the eventual yield of the mission at the $\approx 10$\% level. We note that one of the three major science goals of EVE is to perform mulit-band photometric monitoring of accreting stars within the same fields as the planet survey. Based on the rotation periods, and infrared excess disk fraction, the EVE mission team expects $\sim3$\% of target stars within the youngest fields (e.g. Orion) to exhibit significant accretion signatures (Venuti et al. in-prep). We also note that the 3\,Myr old planet IRAS 04125+2902b was detected in transit about a disk-bearing pre-main sequence star in the Taurus-Auriga star-forming region \citep{2024Natur.635..574B}, and that stars with infrared excess are not automatically precluded from transiting planet detections. 

\subsection{Description of input planet yield models}\label{sec:modelpopulation}

\begin{figure*}
    \centering
    \includegraphics[width=1\linewidth]{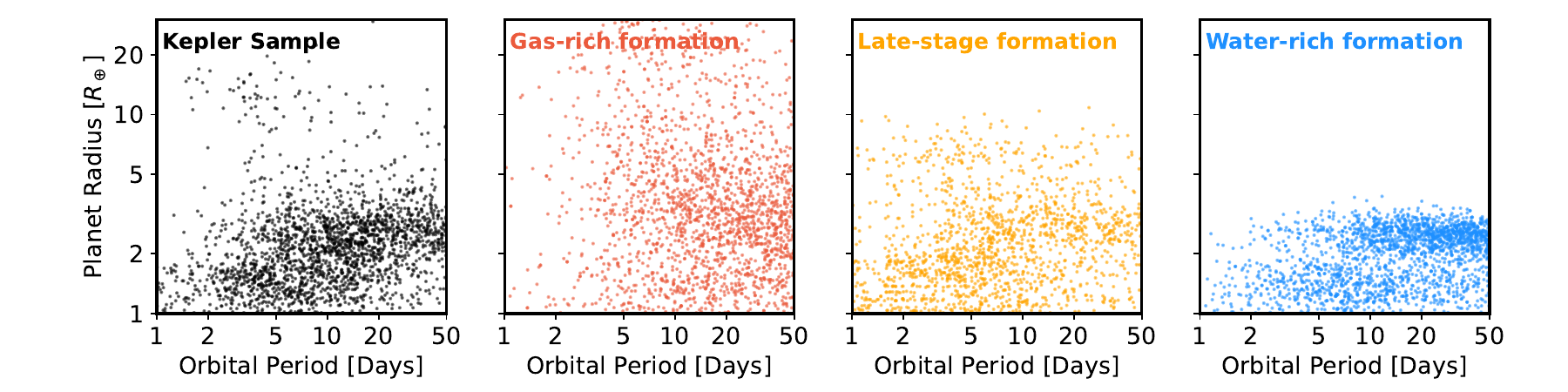}
    \caption{The input sample of the modeled planet population for our mission simulations. These models show the underlying planet population drawn from evolution models from \citet{Rogers2025b} and \citet{2022ApJ...941..186L}, for the gas-dwarf (red), late-stage gas accretion (orange), and water-world (blue) scenarios. Each randomly drawn planet has been evolved to correspond with the ages of their parent stars. No selection biases have been applied to the plotted distributions. Note that the simulated planet populations do not include true gas giants by design, and none of the simulated planets will form eventual hot Jupiters and warm Jupiters. The \emph{Kepler} planet distribution is plotted for comparison. }
    \label{fig:planetinput}
\end{figure*}

To perform exoplanet yield estimates, we use the evolution models from \citet{Rogers2025b} and \citet{2022ApJ...941..186L} to create multiple synthetic populations of planets around our $\approx20,000$ target stars. These incorporate planet evolution tracks for gas-dwarf, late gas-poor formation, and water-rich formation models. These are described in Sections~\ref{sec:rogerssims} and \ref{sec:leesims}, respectively. Each of these models is designed to generate a final population of small planets ($<4\,R_\oplus$) that resembles the super-Earths and sub-Neptunes from \emph{Kepler}. These models do not include true hot Jupiters or warm Jupiters (though they will be easily detectable if transiting), which are likely to have evolved via significantly different migration and formation pathways. The input planet populations for the gas-dwarf, late-forming, and the water-world models, evolved to the ages of their respective host stars, are shown in Figure~\ref{fig:planetinput}. 

For comparison, we also include simulations based on the observed \emph{Kepler} planet population as per the planet frequencies measured by \citet{Kunimoto2020}, without any assumed planet evolution. The \emph{Kepler} occurrence rate simulation incorporates multi-planet system statistics, with a multiplicity rate of 50\% \citep[e.g.][]{2018ApJ...860..101Z,Sandford2019}, with mutual inclinations defined by a $\beta$ distribution of $\alpha=1.5$ and $\beta=5.0$ to replicate the low mutual inclination of \emph{Kepler} systems \citep[e.g.][]{WinnFabrycky2015}. 

\subsubsection{Gas- and water-rich planet evolution tracks}\label{sec:rogerssims}
The first sets of synthetic planets populations are generated using models from \citet{Rogers2025b} for hydrogen/helium dominated gas-dwarfs and water-rich planet evolution models. Under the gas-dwarf model, the radius valley emerges due to the presence, or lack thereof, of hydrogen-dominated envelopes atop Earth-like interiors. Within this model, the dichotomy exists due to atmospheric stripping during protoplanetary disk dispersal \citep[e.g.][]{Owen2016b,Ginzburg2016,Rogers2024a}, and/or atmospheric stripping in the post-disk phase via photoevaporative or core-powered mass loss \citep[e.g.][]{Owen2013,Lopez2013,Gupta2019}. Envelopes are assumed to be of Solar metallicity and are modeled with a mean molecular weight of $2.35$~amu, which produces inflated progenitor planets that can be $\approx 10 R_\oplus$ or larger at young ($\leq$ 30 Myr) ages.

Under the \citet{Rogers2025b} water-world model, the radius valley emerges from distinct formation channels for super-Earths and sub-Neptunes. The former have Earth-like interiors and negligible envelopes, while the sub-Neptunes have rocky cores with $\approx 50\%$ water mass fractions. Their gaseous envelopes are steam-dominated with mean molecular weights of $\approx 18$~amu. As a result, their radii do not significantly contract with time, meaning they are not typically observed at large radii ($\gtrsim 4R_\oplus$) even when young.

Both models' simulated populations are constrained to reproduce the distribution of small, mature (1-10 Gyr) planets detected by \emph{Kepler} via choices in distributions for planet properties such as core masses, initial envelope masses, orbital period and core composition. Then, these constrained models are used to make predictions of planet transit radius as a function of time \citep{Rogers2025b}. For simplicity, we choose to model planets in the post-disk epoch: as such we do not model processes such as core formation or gas accretion as in \citet{Burn2024}. We also do not include any orbital migration in the post-disk phase. In \citet{Rogers:2021}, the initial envelope mass fraction distribution, core mass distribution, mass loss rate efficiency, are derived via forward modeling such that the evolved planet distribution matches the \emph{Kepler} population \citep[e.g.][]{2015ApJ...809....8B}. We adopt this model and randomly evolve 0.7 planets for each EVE target star to the literature age of the star, and adopt the mass, radius, and orbital parameters as our input gas-dwarf and water-dwarf planet population.

\subsubsection{Late-stage formation evolution tracks}\label{sec:leesims}
We also incorporate evolution models from \citet{2022ApJ...941..186L} for planets formed in gas-poor disks. In this set of scenarios, planets with gaseous envelopes in the $2-4\,R_\oplus$ radius range form during late-stage gas accretion in a gas depleted disk \citep{Lee2014,2021ApJ...908...32L}. In the late stages of disk dispersion, eccentricity dampening become less effective and orbit crossings of planet cores begin to become possible. Cores begin to regularly assemble and become massive enough for significant gas accretion to occur. 
Cores more massive than $\approx1-2\,M_\oplus$ can accrete sufficient amounts of gas to appear as $2-4\,R_\oplus$ sub-Neptunes. Lighter cores quickly gather an isothermal maximal envelope and halt their gas accretion. For such small cores, the maximum gas envelope is so small that they will appear as Earths and super-Earth sized planets.

We adopt formation and evolution tracks from \citet{2022ApJ...941..186L}. These models provide the expected envelope mass for each underlying core mass from gas accretion physics as a function of stellar mass and orbital semi-major axis. They further provide the subsequent thermal evolution of planetary radius. The models also allow us to explore the effects of  the assumed level of gas depletion of the disk from which the planets were born (as a function of depletion relative to a Minimum Mass Extrasolar Nebula, MMEN as per \citealt{2013MNRAS.431.3444C}, by 3 and 4 orders of magnitude), and with the effects of photoevaporation optionally incorporated. As default, we adopt models assuming a gas depletion at the $10^{-3}$ level and account for photoevaporation.

We force each model scenario to reproduce the \emph{Kepler} mature aged planet statistics. We draw random mature planets from the \emph{Kepler} occurrence rates as per \citet{Kunimoto2020}. For each random mature planet, we find the nearest model track that replicates its properties. Free variables for the evolution tracks are planet core mass, orbital semi-major axis, initial envelope mass fraction, and stellar mass. We interpolate backward along this track to determine the expected radius of that planet at the age of the young target star sampled by our mission. Note this is different to the approach in \citet{2022ApJ...941..186L}, which initializes their population with sets of core mass distributions, then evolve their planets forward. As such, our synthetic population does not exactly replicate the final population from \citet{2022ApJ...941..186L}.

\section{Transit detection simulations \label{sec:transit_detection}}

We estimate the mission's planet yields based on the hypotheses presented in Section~\ref{sec:modelpopulation}, plus the steady-state \emph{Kepler} planet distribution. We perform the mission yields estimate via a set of forward models that incorporate our best estimates for the photometric precision of EVE in the optical and NIR, and calculate the detectability probabilities based on transit signal-to-noise estimates and injection and recovery results. We run 10 realizations to estimate the yield uncertainties. 

\begin{figure*}
    \centering
    \includegraphics[width=0.7\linewidth]{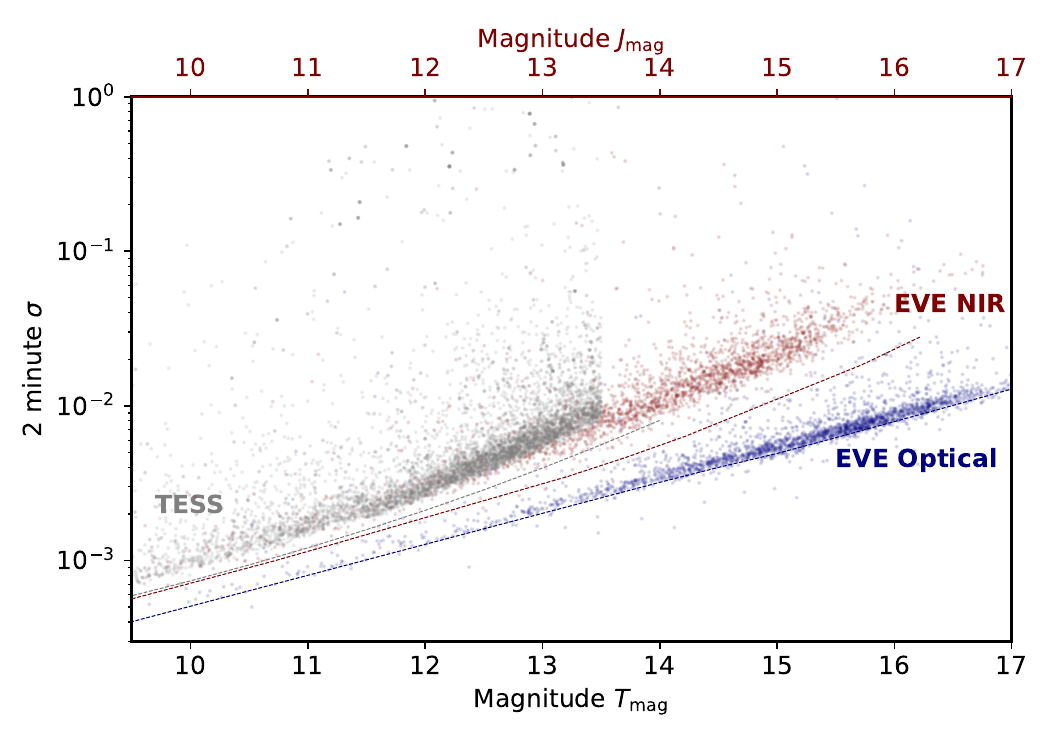}
    \caption{The modeled photometric scatter of EVE in the optical (blue) and NIR (maroon), compared to measured photometric scatter of young stars from \emph{TESS} (grey). The photometric instrument floor for each are marked by the respective dashed lines}
    \label{fig:rms}
\end{figure*}

The predicted photometric floor of the 20\,cm optical and 18\,cm NIR telescopes are shown as dashed lines in Figure~\ref{fig:rms}. The photometric precision floor is determined from an instrument model that accounts for the effective area, throughput, pointing stability, sky and detector background contributions.

Many stars exhibit significant chaotic variability due to processes from accretion, occultation of dipper sources, and flares that cannot be detrended successfully. We do not expect the observed young stars light curve error to reach the photometric precision floor from the engineering model. To capture the effect that a fraction of young stars yield poorly detrended light curves, the simulated photometric precision of an EVE light curve of each target star is drawn from a random distribution based on \emph{TESS} young star light curves. We compare the measured scatter $\sigma_\mathrm{observed}$ of \emph{TESS} light curves from \citet{2024AJ....167..210V} to the predicted photometric floor of \emph{TESS} $\sigma_\mathrm{model}$ from \citet{sullivan15}. The \emph{TESS} light curves have been flattened, with flares $\sigma$-clipped. The distribution of ratios $\sigma_\mathrm{observed} / \sigma_\mathrm{model}$ for young stars in the \citet{2024AJ....167..210V} sample is shown in Figure~\ref{fig:ONC}. We also compute a distribution of $\sigma_\mathrm{observed}$ from light curves of the Orion association to account for the young age of the association. This is detailed in Section~\ref{sec:ONC}. 

The distribution of $\sigma_\mathrm{observed} / \sigma_\mathrm{model}$ is binned into stars specifically from the Orion star forming region, stars $<30$ Myr old, $30-50$ Myr old, and $>50$ Myr old. For each EVE target, the scalar $\sigma_\mathrm{observed} / \sigma_\mathrm{model}$ is drawn from these respective distributions according to the age and membership of the target star. 

Stars with $\sigma_\mathrm{observed} / \sigma_\mathrm{model}>>1$ exhibit significant chaotic variability that does not detrend well for planet search purposes. For each simulated EVE target star, we compute a $\sigma_\mathrm{model}$ based on the instrument model, then randomly draw a scalar factor from the $\sigma_\mathrm{observed} / \sigma_\mathrm{model}$ distribution to determine the final $\sigma$ of the target star. Figure~\ref{fig:rms} shows the simulated photometric precision for EVE target stars in the optical and infrared.

We then perform a forward modeling exercise by drawing planet(s) from our input model planet sample. For each input system, we compute the probability of transit assuming circular orbit, and adopt a uniform impact parameter distribution. For each planet that transits, we compute the transit signal to noise ratio (SNR), with the photometric uncertainty determined from the relationship above.

Specifically, we follow the approach from \citet{2024AJ....167..210V} that maps the SNR of a transit to a detection probability, as a function of stellar type and age. 

\citet{2024AJ....167..210V} characterized this SNR to detection probability mapping via an injection and recovery exercise into the light curves of young stars from \emph{TESS}. In summary, the recovery probability distribution is characterized via a per-star transit signal injection and recovery exercise that is performed over a 1000-point grid sampling planet periods and radii from log-uniform distributions from 0.5 to 30 days, and 1-10$\,\mathrm{R_\oplus}$, impact parameters drawn from a uniform distribution between 0-0.8 (ignoring grazing transits), and transit epochs sampled from a uniformly distributed random phase. 

\begin{figure*}
    \centering
    \includegraphics[width=0.8\linewidth]{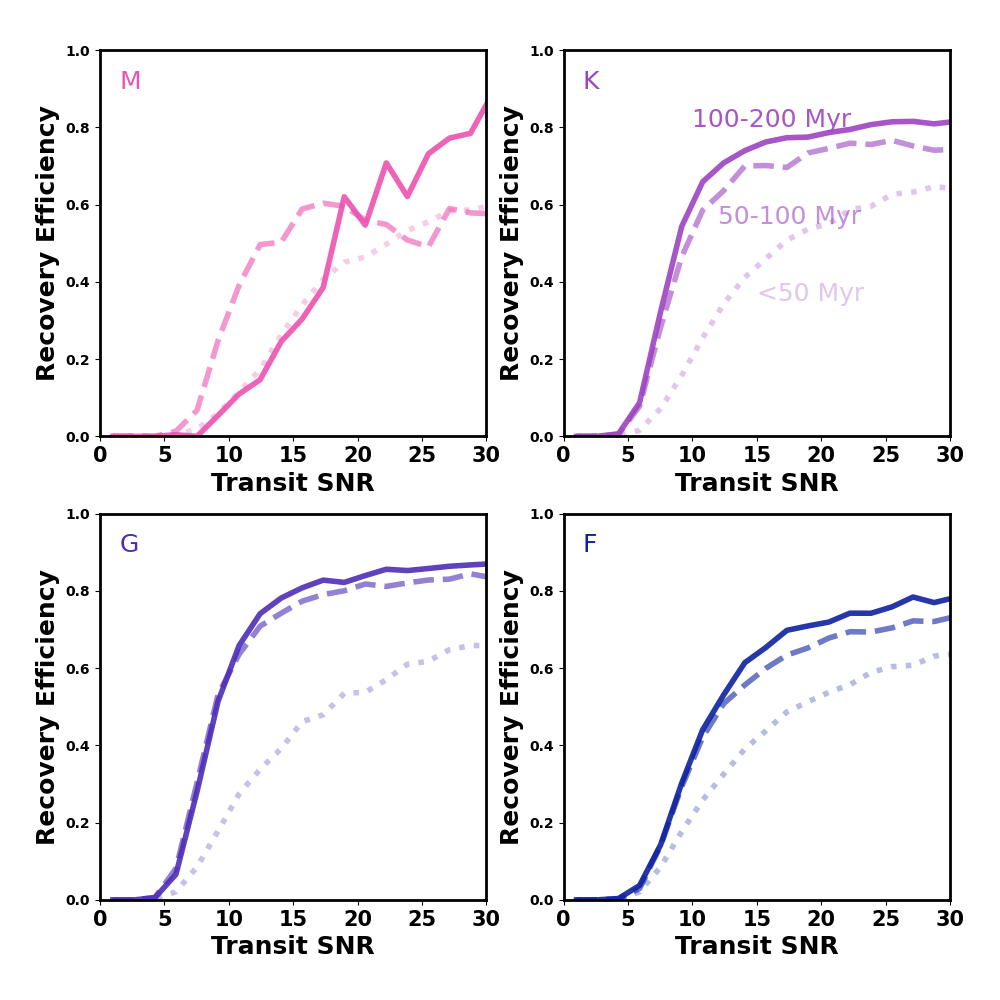}
    \caption{Planet recovery probability given transit signal to noise. Each line shows the probability of recovering a transiting system as determined from a set of young stellar injection and recovery exercises from \citet{2024AJ....167..210V}. The recovery probabilities are characterized based on spectral type and age (dotted for $<50$ Myr, dashed for $50-100$ Myr, and solid for $100-200$ Myr) to incorporate the variability and flaring characteristics that depend strongly on stellar properties for young and adolescent stars. Note the behavior of the recovery probability for M-types stars is qualitatively different from F/G/K stars due to them taking longer to reaching zero-age main-sequence, and exhibiting significant flare activity for $\approx 100$ Myr.  }
    \label{fig:snrvsrecovery}
\end{figure*}

An injected transit is `recovered' if it passes a detection threshold of $8\sigma$ at the correct ephemeris, as per ephemeris matching criterion in \citet{2014AJ....147..119C}. Injected transits are modeled as per \citet{MandelAgol:2002}, with quadratic limb darkening parameters corresponding to a Sun-like star in the \emph{TESS} band of $u_1=0.35$ and $u_2=0.25$. 

The recoverability of an injected signal is quantified based on the theoretical SNR of the transit. For a light curve of 2-minute cadence point-to-point scatter $\sigma_\star$, and an injected transit with depth $\delta$, where $N_{tr}$ transits were observed and the average number of points in transit is $N_{in}$, then the transit SNR is 
\begin{equation}
    \mathrm{SNR_{\star}} = \delta\sqrt{\sum \frac{N_{tr}N_{in}}{\sigma_\star^2}}\,.
\end{equation}

In this work, we adopt the recovery probability distributions from \cite{2024AJ....167..210V}, derived as a function of stellar type and age  (Figure~\ref{fig:snrvsrecovery}), to account for the dependence of stellar variability on these two properties. That is, a young M-dwarf that exhibits significant flare activity should yield a lower transit recovery probability than a similar aged G-star that is expected to exhibit fewer flares, and as such yielding a higher transit recovery probability at the same photometric precision.

In our simulations, for each target star, we first draw the planet properties as described in Section~\ref{sec:modelpopulation} for both the gas-dwarf and the water-world hypotheses. The transit probability is drawn from $\mathrm{Prob} = R_\star/a$, and the impact parameter of the transit is drawn uniformly from $b=0$ to 1 for each transiting system. 
For each transiting planet, we compute its theoretical SNR, adopting the photometric precision curve in Figure~\ref{fig:rms} for $\sigma_\star$.

We also account for contamination from adjacent stars in crowded fields. In crowded fields, unresolved or partially resolved neighboring stars dilute the transit depths, reducing the transit SNR and the detection probability. We assume a pixel response function of 10\arcsec{} for the optical and NIR telescopes. The angular stellar density in each field is estimated via a GAIA query that estimates the average stellar density over the $5^\circ\times5^\circ$ target field of view. The total light is summed over a circular aperture of radius 10\arcsec, and the planetary transit depth is diluted accordingly.

\subsection{Accounting for the Orion stellar sample}
\label{sec:ONC}

A substantial fraction of young planets detected by EVE are expected to be from the Orion star forming region. These planets will be younger than most transiting planets known today.  To examine the robustness of our simulations for this population of stars, we perform a photometric noise characterisation of stars from Orion to confirm the noise assumptions we made in the transit signal to noise calculations remain valid for such young stars.

\begin{figure}
    \centering
    \includegraphics[width=1\linewidth]{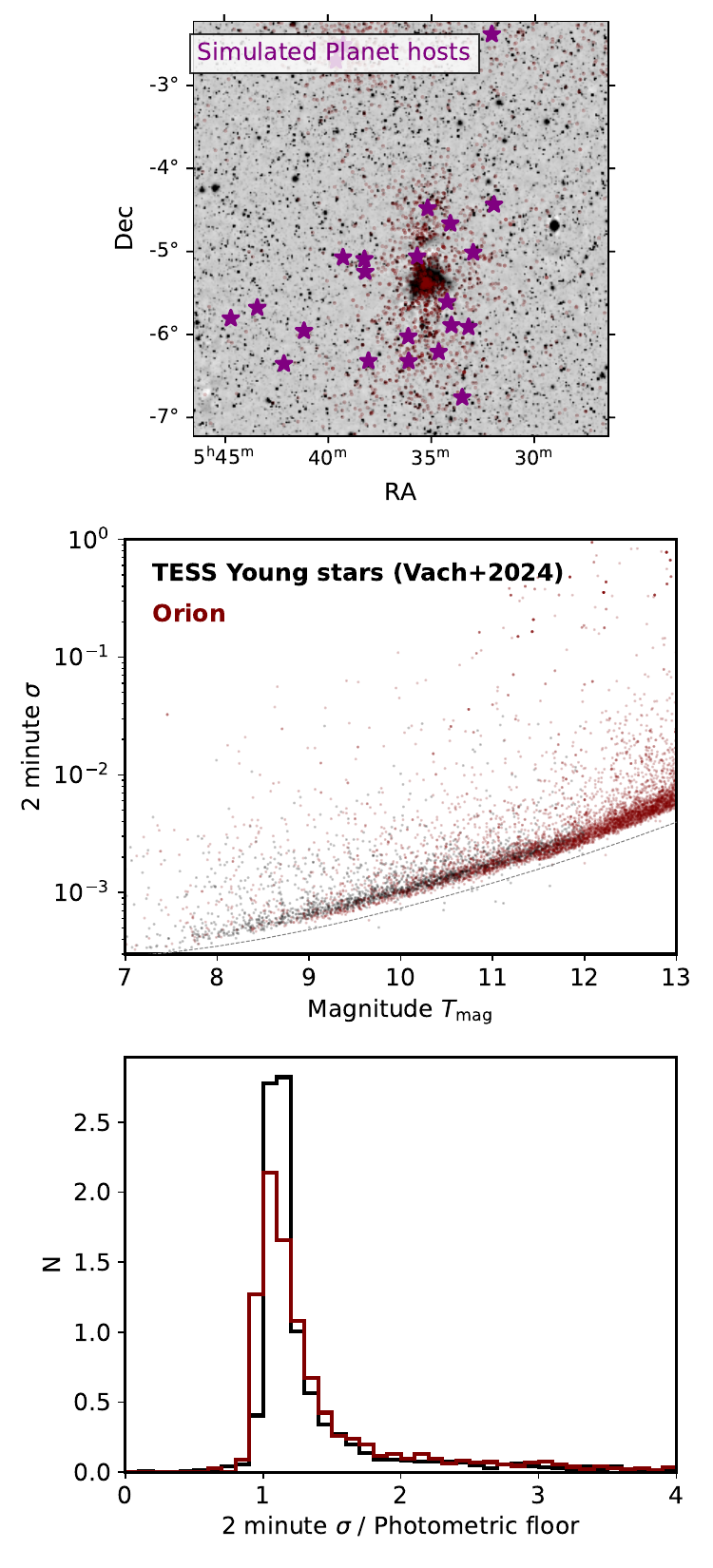}
    \caption{\textbf{Top} One of the Orion target fields is shown in 2MASS $J$ band. Stars with simulated planets recovered from EVE are marked in purple to illustrate the expected spatial distribution of Orion planets. \textbf{Middle} The \emph{TESS} light curve scatter for young stars from \citet{2024AJ....167..210V}, compared to those from Orion members. \textbf{Bottom} The photometric scatter distribution, as a function of deviation from the photometric noise floor $\sigma_\mathrm{observed} / \sigma_\mathrm{model}$.}
    \label{fig:ONC}
\end{figure}

The Orion membership catalog from \citet{2025A&A...702A..63R} is plotted spatially in Figure~\ref{fig:ONC} over one example $5^\circ\times5^\circ$ field of Orion EVE field. We retrieve all available \emph{TESS} full frame image light curves \citep{qlp2020a,qlp2020b} for Orion members from \citet{2025A&A...702A..63R}, and detrend the light curves via the \textsc{keplerspline} routine \citep{Vanderburg:2019}. We make use of the same detrending settings as per \citet{2024AJ....167..210V}, on which our planet retrieval exercises are based. Note that we perform our analysis using Full Frame Image light curves due to their greater availability at the magnitude range of the Orion members. Where necessary, we modify all reported light curve scatter root-mean-square (RMS) values to 30 minute bins (for example, light curve scatter measured from 30 minute cadence light curves are corrected by a factor of $\sqrt{15}$ to be comparable to 2 minute light curves).

Figure~\ref{fig:ONC} shows the RMS of the light curves from Orion members. For any EVE target star that is part of the Orion star forming region, we adopt this Orion specific RMS distribution, instead of one prescribed by the general young \emph{TESS} population.

Notably, stars from the Orion Nebula Core exhibit significantly higher photometric scatter than those in the greater Orion association. This is likely due to the significant background from nebulosity contributing to the photometric noise. As a result, we do not expect many planets to be found within the Orion Nebula Core itself from our yield estimates ($\sim 0.4$ planets are expected in the nebula core from the mission), and the number of stars within the core that are amenable for planet search $(\sim 1000)$ is small enough to not significantly impact our occurrence rate calculations.

\subsection{Impact of simultaneous optical and NIR bands for transit recovery}
\label{sec:OpticalNIR}

Young planet surveys benefit significantly from simultaneous optical/NIR observations. Unlike traditional all-sky and magnitude limited surveys that can be biased towards F, G, and K stars, we need to sample as many stars in a single association as possible to achieve the maximum planet yield. As such, M-dwarfs dominate the input sample. 

\begin{figure}
    \centering
    \includegraphics[width=1\linewidth]{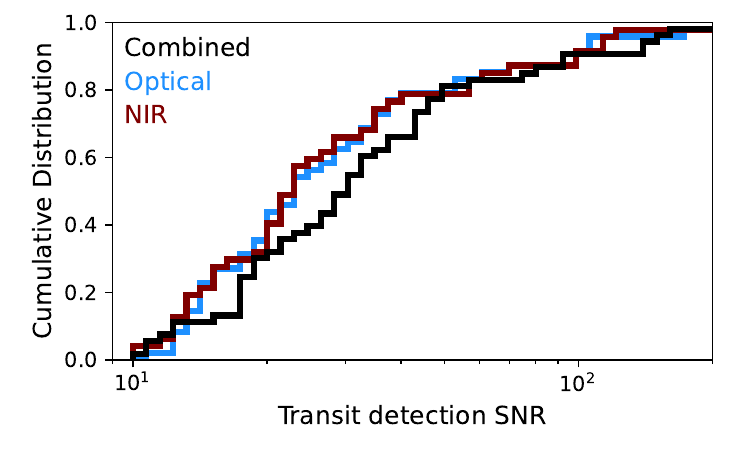}
    \caption{The optical and NIR channels make nearly equal contributions to the total planet yield of the mission. The cumulative histogram shows the transit signal-to-noise of detected planets in the optical (blue), NIR (red) channels, and in the joint analysis (black).}
    \label{fig:opticalvsNIR}
\end{figure}

In addition, the NIR band captures a significant fraction of the total flux from a young star. Figure~\ref{fig:opticalvsNIR} shows the flux and transit signal-to-noise contribution of the NIR channel compared to the optical channel. A $1\,M_\odot$ star at 10\,Myr has an effective temperature of $T_\mathrm{eff} = 4400$\,K, and its energy flux over a joint $J$ and $H$ band is 0.95 times that of its emissions over the $TESS$ optical band. A $0.5\,M_\odot$ star at the same age has an effective temperature of just $3700$\,K, and emits more flux over the $J+H$ band than the optical \emph{TESS} band by a factor of 1.4. On average, the infrared channel contributes to $\approx 40$\% of the total flux of stars suitable for EVE's planet search program. As such, flux contributions from a simultaneous NIR band provide a significant improvement over optical-only observations of star-forming regions. 

The photometric precision in the NIR band is computed via a similar method to that of the optical band. We determine the NIR band flux for each target star (simulated to be the summed flux over $J$ and $H$), and convert that flux to a photometric precision via the instrument model, adopting the same $\sigma_\mathrm{obs} / \sigma_\mathrm{model}$ distribution as that in the optical. We note that NIR background noise is not accounted for in this modeling, which will have an adverse impact on the photometric precision for the faintest targets. The total transit SNR of a simultaneous optical + NIR system is computed as the quadrature addition of the transit SNR in each channel. This is similar, but not equivalent to, adding the flux of the two channels before computing transit signal-to-noise due to non-photon limited noise sources that are accounted for in the stellar flux - photometric precision relationship. Figure~\ref{fig:opticalvsNIR} shows the contribution of the NIR band to the transit detection signal in our simulations as compared to that of the optical transits. The NIR channel contributes on average 44\% of the total signal-to-noise of a transit detection making it nearly equally important to the total planet yield of the mission. 

Using the yields from the gas-dwarf models of \citet{Rogers2025b} as an example, we find an expected yield of $99\pm7$ planets younger than $50$\,Myr over the primary mission from simultaneous optical and NIR observations, with $\approx 20,000$ target stars observed over 30 fields for least 30\,days per field. In comparison, if only optical channel observations are conducted, we would expect only $97\pm4$ planets during the primary mission.

\subsection{NUV correction of stellar activity improvements to transit recovery}
\label{sec:NUVcorrection}

Two sequential flares can masquerade as planetary transits as a result of significant light curve whitening and detrending. Masking of light curve regions affected by flares can improve the false alarm probability in transit searches, though the flare frequency of many pre-main sequence M-dwarfs \citep[e.g.][]{2022AJ....163..147G,2022AJ....163..156M} makes it impractical to mask all affected light curve regions. 

\citet{2025AJ....169...27H} demonstrated that the optical light curves of flares can be accurately reconstructed from simultaneous NUV photometry. Optical flares are generated from chromospheric heating that occurs at a slower timescale than the prompt UV emission detected in the NUV channel of EVE. \citet{2025AJ....169...27H} observed simultaneous \emph{Swift} and \emph{TESS} flare observations, and showed that the shape and peak of the optical light curve can be modeled from the time integral of the NUV event. 

Transit injections and recoveries were performed, demonstrating that simultaneous modeling of injected transits in \emph{TESS}, after flare modeling and removal from UV observations, can significantly increase transit detection sensitivity, especially for the smallest planets \citep[See][Figure 9.]{2025AJ....169...27H}. We repeated the planet yields simulation, accounting for the availability of NUV-dependent flare modeling in whitening the optical and NIR light curves pre-transit detection. Since our detection probability is mapped as a function of the SNR of a transit, the age of the star, and the spectral type of the star (Figure~\ref{fig:snrvsrecovery}), we simply adopt the detection probability functions for an older, non-flare dominated stage of each spectral type for an optimistic planet yields estimate. 

To estimate the yield in this scenario, we adopt the transit SNR-recovery curve of a 100\,Myr G-type star for all target stars that can be detected in quiescence in the NUV. Only stars detectable in quiescence in the NUV channel, with a flux of $>5\times10^{-13}\,\mathrm{erg\,s}^{-1}\,\mathrm{cm}^{-2}$, are found to benefit from flare modeling and removal in this realization of the yields simulation. At 100 Myr, flares occur infrequently and the photometric variability of a G-star is spot modulation dominated \citep[e.g.][]{2024AJ....168...60F}, and is representative of a best-case scenario flare modeling enabled planet search.  We find that the inclusion of flare modeling and removal for these brightest stars that are visible in the NUV in quiescence will increase the expected gas-dwarf planet yield by $\approx 5\%$.

\section{Results and discussion}

\subsection{Simulated planet detection results}\label{sec:yield}

\begin{figure*}
    \centering
    \includegraphics[width=1\linewidth]{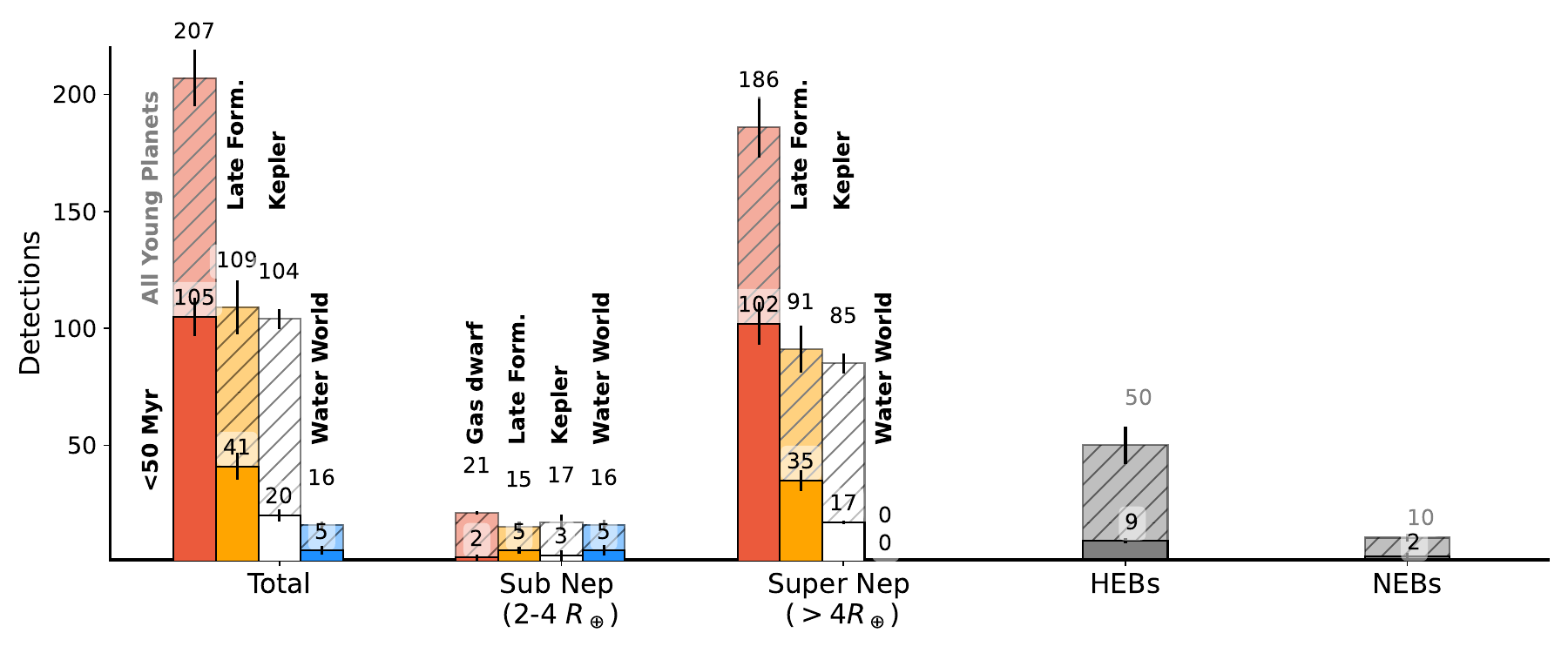}
    \caption{Expected planet yield of a simultaneous NUV, optical, and NIR wide field survey, assuming a 20\,cm aperture optical telescope and 30\,day stare durations per field over a 2.5 year primary mission. Planet yields based on the gas-dwarf hypothesis from \citet{Rogers2025b} are plotted in red, late-stage formation gas-dwarfs in orange \citep{2022ApJ...941..186L}, water-world hypothesis from \citet{Rogers2025b} in blue, and based on steady-state \emph{Kepler} statistics in black. The planet yield for young planets ($<50\,$Myr) are plotted in solid colors, and those of all ages about kinematically associated stars in hashed light colors . A small number of super-Earths ($<2\,R_\oplus$) are detected in some of the simulations, and account for the remaining planets. The expected astrophysical false positive rates are also reported, with all planet-like false positives about 50 Myr or younger stars marked by the solid bars, and those about all young kinematically associated stars in hashed bars. Only hierarchical eclipsing binaries (HEBs), and nearby eclipsing binaries (NEBs) are expected to contribute towards the false positive statistics of EVE.}
    \label{fig:planetyield}
\end{figure*}

Figure~\ref{fig:planetyield} shows the expected yields of our simulations for the gas-dwarfs, late-stage formed gas-dwarfs, water-worlds, and \emph{Kepler} non-evolving planets. These simulations assume a 2.5 year primary mission surveying 30 young stellar fields with a minimum stare time per field of 30 days. 

If sub-Neptunes are predominantly formed with extensive low mean molecular weight gaseous envelopes early in the formation process, then EVE is expected to detect \gasyield{} young planets that have ages $<50$ Myr. The majority of the detectable young planets in this scenario should have radii $>4\,R_\oplus$, but masses $<20\,M_\oplus$, reflecting the progenitor states of the vast majority of sub-Neptunes and super-Earths found around mature aged stars. If gas-dwarfs were formed in a depleted disk, the expected yield would range between 20-40 planets. Forward modeling of the eventual planet yield will help determine when the planets were formed in the gas disk. Section~\ref{sec:discusslatestageformation} explores in-depth different assumptions of the gas-poor formation process and their influence on the detected planet distribution. If instead all planets are formed water-rich then the predicted planet yield is only \wateryield, as most of the expected population fall below the detection threshold of the survey at $\approx 3-4 R_\oplus$. Simulations based on the \emph{Kepler} planet occurrence derived by \citet{Kunimoto2020}, assuming no planet evolution, yields \kepleryield planets from the mission. 

The expected planet population is shown in Figure~\ref{fig:planetdistr}, corresponding to the detectable population of the input planet distribution from Figure~\ref{fig:planetinput}. Most planets are expected to have radii in the sub-Neptune to super-Neptune range, with periods $\leq$30 days as restricted by the per field stare duration of the mission. In addition, we also report the number of planets of all ages about kinematically associated stars that will be detected by the mission. In the gas-rich scenario we expect $198\pm12$ planets of ages $<1$\,Gyr to be identified. 

False positive scenarios addressed in our models include eclipsing binaries (EBs), hierarchical eclipsing binaries (HEBs), and nearby eclipsing binaries (NEBs) with transit depths $<3\%$. The contribution of these false positive scenarios are estimated in Appendix~\ref{sec:falsepos}, and included in Figure~\ref{fig:planetyield} for comparison. No EBs are expected to be identified as planet candidates. A total of $33\pm2$ HEBs and NEBs false positive candidates will exhibit transits of planetary depth about stars younger than 50\,Myr. Using NIR and optical depth differences, we can eliminate a substantial fraction of these false positive candidates. Out of the 42 false positive candidates, 30 can be eliminated due to NIR optical color differences, leaving $12\pm2$ false positive candidates in total. Figure~\ref{fig:planetyield} shows the expected false positive yield for the primary target list of $<50$\,Myr stars, and for all kinematically associated stars residing in the fields to be sampled by EVE. The expected number of planet candidates from the mission will be the \gasyield{} true young planets plus these false positive planet candidates.  

The reported uncertainties are the standard deviations from the 5 replicate runs. We do not account for uncertainties in the population models. Section~\ref{sec:discusslatestageformation} discusses the range of possible planet yield outcomes as a function of model assumptions on gas disk depletion and the effectiveness of photoevaporation. The young planet population we expect to detect has the power to differentiate between these models.

\begin{figure}
    \centering
    \includegraphics[width=1\linewidth]{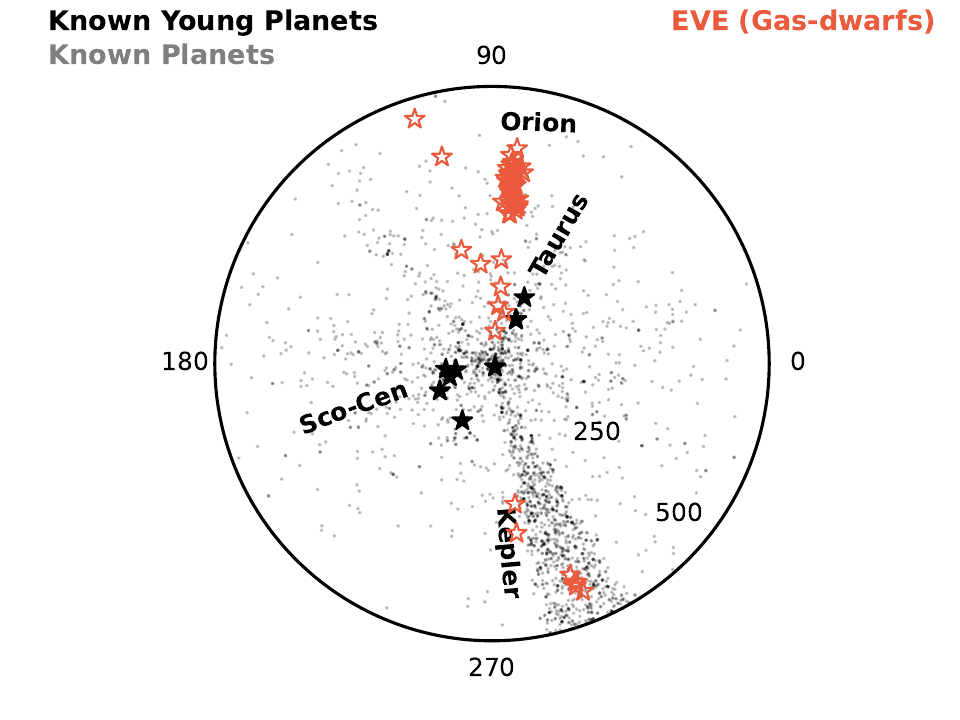}
    \includegraphics[width=1\linewidth]{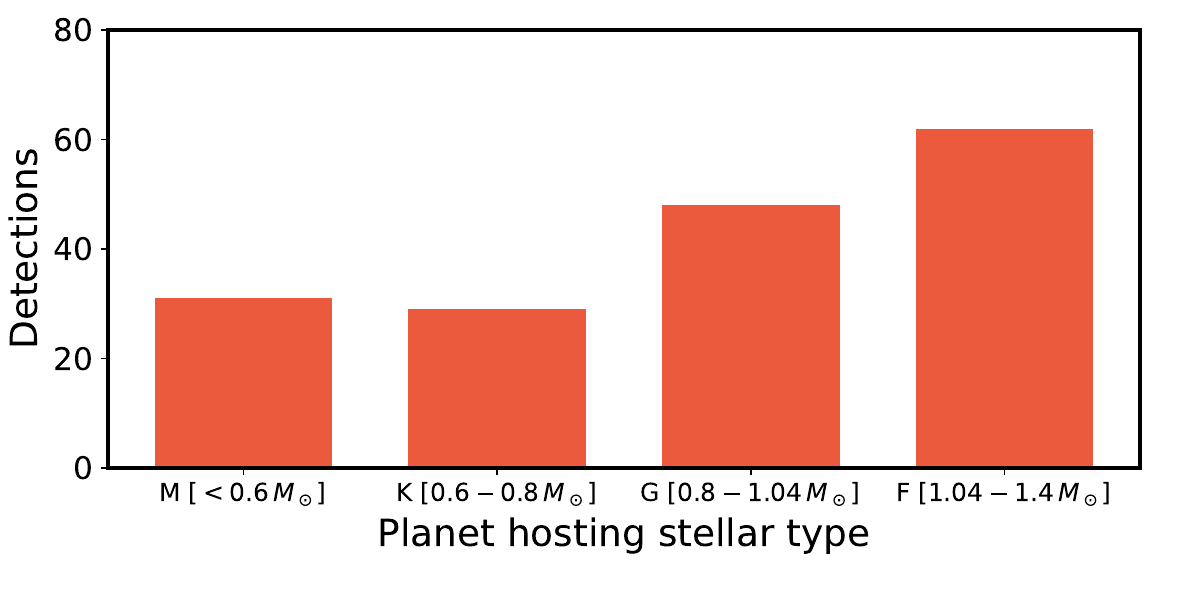}
    \includegraphics[width=1\linewidth]{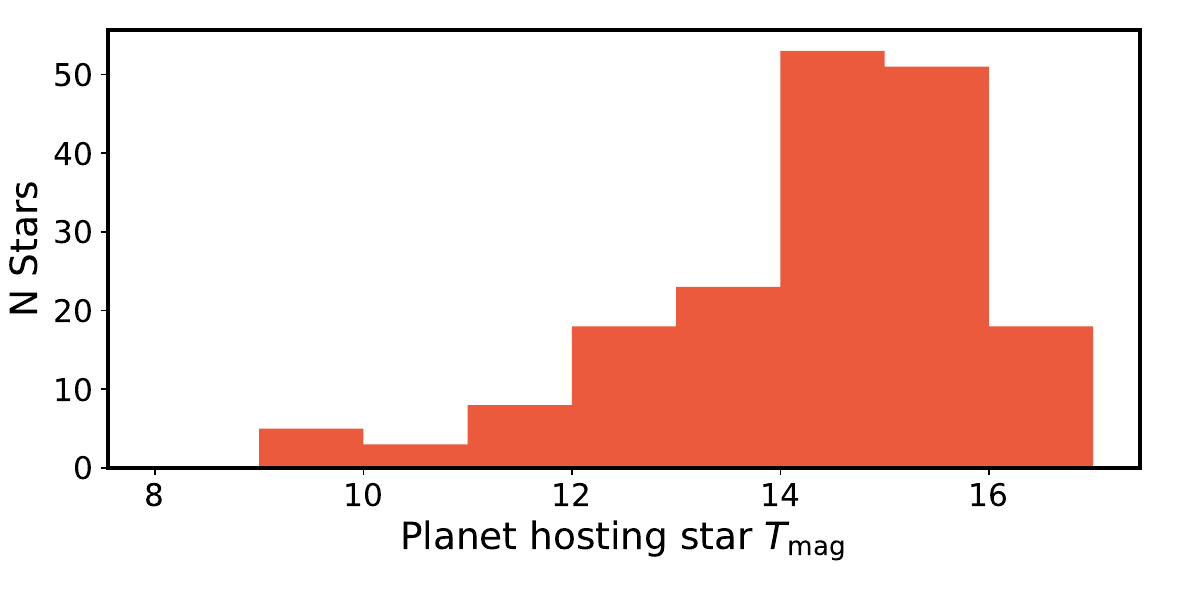}
    \caption{Distribution of expected EVE planets from the gas dwarf model shown with Galactic coordinates, stellar type, and magnitude. The \textbf{Top panell} the existing planet population is plotted in grey, with the known young planets highlighted. The dense star forming regions in Orion are $\approx 300$ pc away, and late-type stars in these regions are too faint to yield significant numbers of planet detections with \emph{TESS}. In contrast, most known systems have been discovered in the close-by Scorpius-Centaurus, Taurus, and $\beta$ Pic regions. The \textbf{Middle and bottom panels} show the expected planet-hosting star spectral type and \emph{TESS} band magnitudes.}
    \label{fig:distances}
\end{figure}

\begin{figure*}
    \centering
    \includegraphics[width=0.9\linewidth]{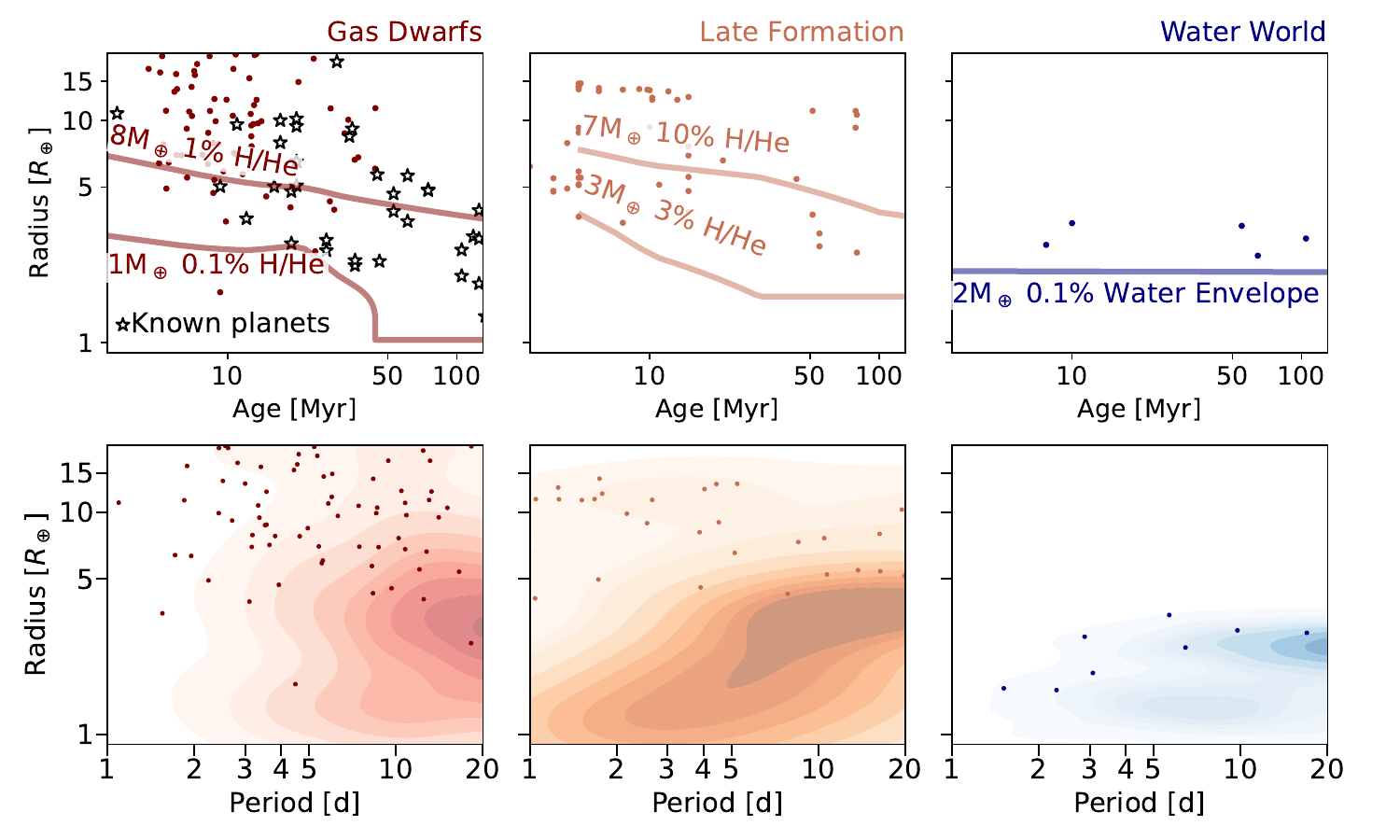}
    \caption{The distribution of simulated planet detections in radius, period, and age. Known young planets are plotted as stars on the top left panel for context. \textbf{Top row:} The radius distribution of planet discoveries as a function of age. The model curves show the modeled radial evolution of select gas-dwarf (red), late-stage formation with $10^{-3}$ disk depletion (orange), and water-world (blue) scenarios as per \citet{Rogers2025b} and \citet{2022ApJ...941..186L}. \textbf{Bottom row:} The period radius distributions, with the detected planets plotted in dots, and the input distribution as 2D histograms.}
    \label{fig:planetdistr}
\end{figure*}

\textbf{EVE will access more distant star forming regions than \emph{TESS}} Previous exoplanet discovery missions were restricted to searching star forming regions within 100\,pc either due to magnitude limits (in the case of \emph{TESS}) or the placement of the stellar clusters on the sky (in the case of \emph{K2}). This limitation has severely restricted the stellar sample that could be surveyed and has thus far prevented the accumulation of a statistically significant and unbiased young planet sample. The mission will have the capability to detect planets in star forming regions more distant than \emph{TESS}, empowering the first meaningful survey of star forming regions in Orion and Vela OB. Figure~\ref{fig:distances} shows the Galactic distribution of previously known young planets from \emph{TESS} and \emph{K2} against those that are expected to be found by the mission. 

Most planet systems detectable from the Orion complex can be followed up with existing ground-based facilities. The magnitude range of expected EVE planet-host stars are shown in Figure~\ref{fig:distances}. Within the Orion fields, 68\% of planet-hosting stars have magnitudes between $T_\mathrm{mag}$ of 11.6 to 15.2. \emph{TESS} and \emph{K2} young planet candidates within this range are routinely followed-up with ground-based photometric and spectroscopic facilities for validation (e.g. K2-33 \citealt{David2016,Mann:2016}, TOI-1227 \citealt{2022AJ....163..156M}, IRAS 04125+2902 \citealt{2024Natur.635..574B}).

Young stellar associations and star forming regions naturally reside in the galactic disk and areas of high stellar density. The highest stellar density fields to be surveyed by EVE reside in Sagittarius near the galactic plane, and have densities of $\sim 3$ stars per $10\arcsec$ PSF. Section~\ref{sec:falsepos} specifically computes the false positive rate due to background eclipsing binaries, reporting a likely low false positive rate of $\sim10\%$ for the total candidate yield from the mission. We note many false positive scenarios can be eliminated from the color-dependence in the transit depths measured in the NIR and optical channels. 

Nevertheless, validation of the planet candidates from EVE will likely involve ground-based meter-class photometric facilities to confirm on-target transits via additional multiple bands \citep[e.g.][]{2018AJ....156..234C,2019AAS...23314005C}, spectroscopic reconnaissance observations to confirm lack of large radial velocity variations indicative of stellar binary systems, and diffraction limited imaging to check for the presence of background or associated stars within seeing-limited spatial scales. This is routinely performed for all planet discoveries from the \emph{TESS} mission by the community today. Note the regions of regard for the mission are equatorial, and the stars will be accessible from a number of large-aperture radial velocities (VLT, Keck, Gemini-North) which help to address follow-up and potential mass measurement concerns.

\textbf{Benefit of multi-band observations} Simultaneous multi-band observations also contributes significantly to the increased planet yield of the mission compared to \emph{TESS}.  With only the optical channel, the mission will yield $\approx 3\times$ the number of young planets as that from \emph{TESS} in the same age range. The addition of the NUV and NIR channels increase that to $\approx 6\times$ the current young planet population; such a mission design is one way to produce a statistically significant, uniform sample of planets in star forming regions.

\textbf{Multi-planet statistics} We also incorporate system multiplicity with the \emph{Kepler}-statistics simulations. Within these simulations, $25\pm 4$\% of the detected planets were the inner planet in a multi-transiting planet system. However, we expect to find only $1.5\pm1.1$ ($\approx 2$\%) multi-transiting planet systems out-right in the primary mission. 
The additional planets in the remaining EVE-detected systems have periods longer than that easily detectable for a 30-day stare campaign.
As such, we would expect multi-planet statistics to significantly improve with additional re-visits of surveyed fields during possible mission extensions.

\textbf{Safe modes} The planet yield will not be significantly affected by short safe mode interruptions to the science observations\footnote{Safe mode simulations assume a random 10\% segment of a 30-day stare sector is lost, and that 3 such safe-mode interruptions occur over the 2.5 year primary mission.}. We extended the simulations in Section~\ref{sec:transit_detection} to simulate random gaps in 1--5 randomly selected fields during the 30 field survey, and found the planet yield to decrease by only 5\% due to these random gaps. 

\subsection{Distinguishing between gas accretion scenarios}
\label{sec:discusslatestageformation}

A significant young planet population is required to determine when gas-rich planets formed in their disks. Both photoevaporation-dominated gas-rich planet formation models and the gas-poor models replicate the present day close-in planet population. These two scenarios will also yield the same final envelope mass fraction so as to replicate the \emph{Kepler} distribution, and so atmospheric follow-up of mature-aged planets ($\gtrsim1$\,Myr) is unlikely to differentiate between these two scenarios.

For our simulated mission, the gas-dwarf models from \citet{Rogers2025b} predict \gasyield{} planets $<50$\,Myr old. This scenario yields the most planets since it assumes the cores of Neptunes and sub-Neptune are massive enough to accrete significant envelopes early on (but not undergo runaway processes to form true giant planets). These envelopes then undergo evaporation and contraction, and significant radial evolution occurs within $\approx 100$ Myr. Alternatively, if the gas envelope was accreted after significant depletion in the gas disk as per \citet{2022ApJ...941..186L}, then the young planet sizes will be significantly smaller, with less radial evolution, and fewer planets will be recovered. In the late formation scenario, the planet occurrence rates should converge to that of the \emph{Kepler} distribution. 

\begin{figure}
    \centering
    \includegraphics[width=0.9\linewidth]{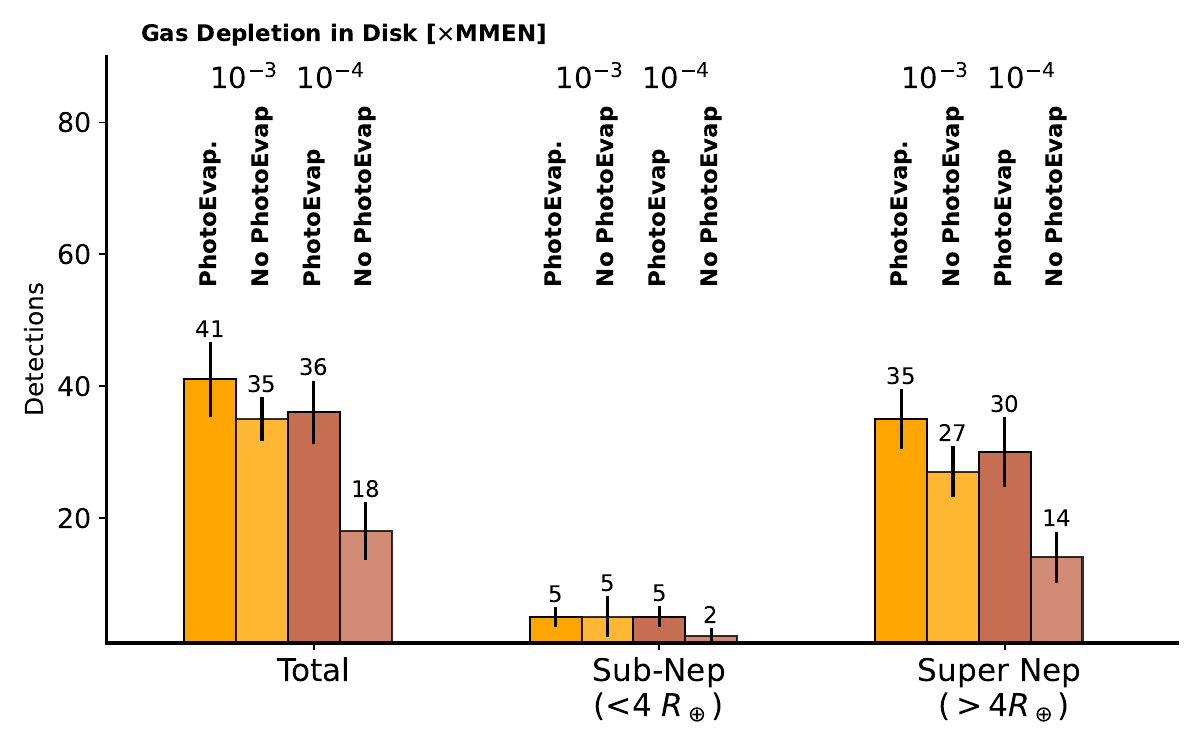} \\
    \caption{Expected yield for late formation scenarios for systems $<50$ Myr old. Planet yields are shown for formation in disks depleted at the $10^{-3}$ and $10^{-4}$ level compared to the MMEN, and with and without account for post formation radial evolution from photoevaporation.}
    \label{fig:lateformYield}
\end{figure}

Figure~\ref{fig:lateformYield} shows the simulated planet yield given different modeling assumptions in the late stage formation scenario. Simulations were performed for planets forming in disks depleted of gas at the $10^{-3}$ and $10^{-4}$ level, with photoevaporation included and excluded to demonstrate its influence on the resulting population. Planets formed when the gas disk is depleted at the $10^{-3}$ level should yield \lateyield{} planets, most of which are of super-Neptune size. If the planets formed later, the number of super-Neptunes with significant gas envelopes naturally declines. 

Note that in our simulations, the input planet population was tuned to reproduce the \emph{Kepler} planet distribution. As such, the inclusion of photoevaporation in the model increases the young planet radii and thereby the predicted planet yields. That is, when the mechanics of mass loss are optionally accounted for in the simulations, the planets naturally need to have formed with more massive envelopes than when mass loss is ignored, in order to reproduce the same final planet distribution. This interpretation differs slightly to \citet{2022ApJ...941..186L}, which compares planet population with the same starting conditions and mass loss optionally incorporated. 

\subsection{Occurrence of newly-formed planets}\label{sec:OR}

\begin{figure}
    \centering
    \textbf{Injected}\\
    \includegraphics[width=0.9\linewidth]{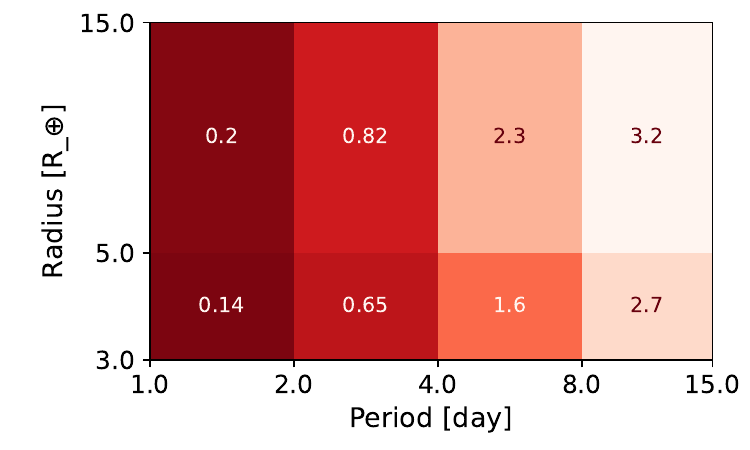} 
    \textbf{Retrieved}\\
    \includegraphics[width=0.9\linewidth]{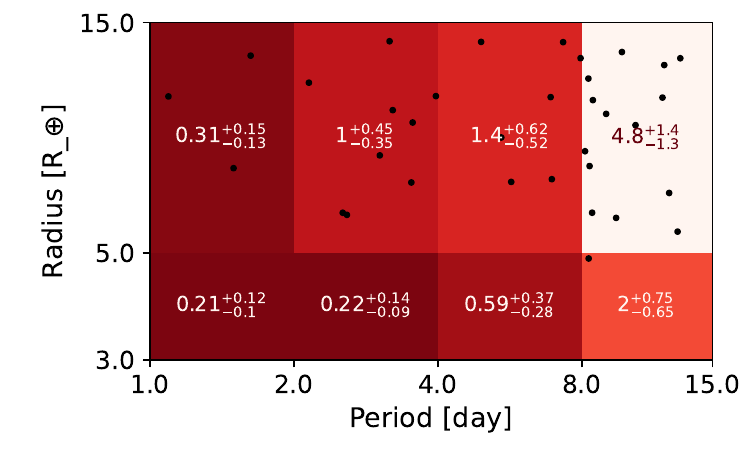} \\
    \caption{The simulated mission can determine the planet occurrence rate of the young planet population. The true injected occurrence rates are shown on top, while recovered occurrence rates via simulated planet detections are shown on the bottom. The color scale corresponds to the planet occurrence rate (injected for top plot, retrieved for bottom plot). }
    \label{fig:OR}
\end{figure}

To estimate the occurrence precision such a mission can attain, we perform a population Monte Carlo Approximate Bayesian Computing (ABC) of the simulated gas-dwarf planet population as per \citet{Beaumont:2009,cosmoabc,2018AJ....155..205H,2019AJ....158..109H,Kunimoto2020}. Briefly, ABC occurrence rate modeling is well suited towards small planet yield studies and was originally developed to better estimate the frequency of Earth-like planets $(\eta_{\oplus})$ from \emph{Kepler}. ABC requires a forward model and a distance function. The forward model is a function that tests the probability of recovering a planet given 1) transit probability, 2) detection pipeline efficiency, and 3) false positive probability. In particular, \citet{Kunimoto2020} notes that, unlike traditional techniques that average the detection efficiency of planets over a range of stars to reduce the dependence of detection probability to planet period and radius, ABC instead accounts for these detection efficiencies on a per-system basis. A distance function makes the comparison between the forward model results and the observation in radius, period, and spectral type, such that the results can be directly comparable to that from \emph{Kepler} and \emph{TESS} despite the differences in the sampled stellar population. 

The occurrence rate from this mission will be determined via a survey of $\approx 20,000$ stars with ages $<50$ Myr in star forming regions over the course of its primary mission. For comparison, \citet{2024AJ....167..210V} identified $7219$ stars younger than $200$ Myr that were suitable for planet search in the first five years of \emph{TESS}, while \citet{2025AJ....169..208F} made use of 1374 stars for their occurrence rate calculations.

The recovered occurrence rate maps are shown in Figure~\ref{fig:OR}. We retrieve a planet occurrence rate of $11.0_{-1.6}^{+1.8}$\% within the bounds of $3<R_p<15\,R_\oplus$ and $1<P<15$\,days, exactly in agreement with the injected planet frequency of 11.5\%.

\subsection{Preliminary atmospheric characterization via multi-band photometry}

Precision transit observations of planets in multiple bands from such a space observatory can also characterize the atmospheres of the planet upon discovery. 
Chromatic transit depth differences may be expected due to Rayleigh scattering, haze, and/or dust in a atmosphere with a sufficiently large scale height \citep{2017MNRAS.468.3418G}. Young, inflated transiting planets, especially those with low mean-molecular weight envelopes, are expected to have low surface gravity and thus a relatively large scale height. 

Those young planets that have been observed thus far by HST and JWST have subsequently exhibited strong spectral features \citep[e.g.][]{Barat:2024a,Thao:2024,Barat:2024b}. In particular, HST observations of the 10\,Myr planet K2-33b by \citet{Thao:2023} found the optical transit of the planet to be $2\times$ deeper than a corresponding NIR transit. Possible interpretations for this discrepancy include strong absorption by photochemical haze in the upper atmosphere, or the presence of dusty, circumplanetary rings  \citep{Ohno:2022}. 

\begin{figure}
    \centering
    \includegraphics[width=0.4\textwidth]{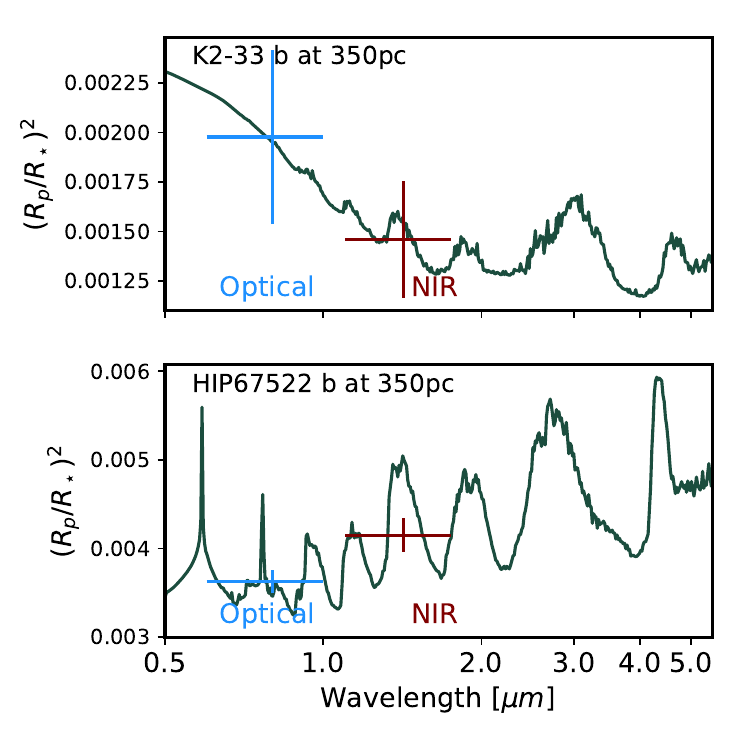}
    \caption{EVE will probe transit depth differences in the optical and infrared. Spectral models of K2-33\,b and HIP 67522\,b are shown, and transit depths and uncertainties are computed over the EVE bands over these models assuming a distance of 350\,pc -- a characteristic distance for regions targeted in Figure~\ref{fig:maps}. The K2-33\,b model is adopted from \citet{Thao:2023}, showing absorption due to Tholin haze. The HIP 67522\,b model for a clear and non-star-spot affected atmosphere is adopted from \citet{Thao:2024}.}
    \label{fig:atmosphere}
\end{figure}

The EVE mission concept has the potential to identify candidates that exhibit strong differences between optical and infrared transit depths, identifying compelling candidates for further atmospheric follow-up, i.e by JWST. Figure~\ref{fig:atmosphere} shows the spectral models of planets equivalent to K2-33\,b and HIP 67522 b, propagated to a distance of 350\,pc (in line with most fields that such a mission will survey) with the averaged transit depths and uncertainties over the multi-band channels marked. These observations have the potential to reveal tentative atmospheric features to differentiate between clear and hazy atmospheres, and reveal planets that are experiencing significant hydrodynamic escape. Such planets would be prime targets for JWST follow-up to better understand the escape process early in the evolution process of Neptunes and sub-Neptunes. 

\section{Conclusions}

We showed through these simulations that a 20\,cm class mission within the scope of the NASA SMEX Program that performs a multi-band photometric survey of select star forming regions from low Earth orbit can significantly increase the statistical sample of newly formed planets known. These planets will be among the youngest transiting planets discovered to date. Many of the targeted regions are too distant, and their late-type stars too faint, for \emph{TESS} and other 0.1\,m class facilities to yield a significant number of planet detections.

Small planets orbiting close-in may form with substantial hydrogen/helium rich envelopes. Depending on when they accrete their gas envelopes, these planets may undergo boil off and photoevaporation and shrink dramatically during the first hundred million years of evolution. Alternatively, the small planets are born with heavy water-rich envelopes with radii that change little over their lifetimes. These models are degenerate in planet mass and radius at the mature ages that have been surveyed by \emph{Kepler} and \emph{TESS}, limiting our ability to understand the formation and evolution of such planets. At planet ages $<$100 Myr, however, the models diverge significantly with the gas-dwarf planets exhibiting radii up to 4$\times$ larger than the water-worlds. Only 20 transiting planets are known today with ages $<50$ Myr, and they are insufficient in differentiating between these models today. 

To demonstrate the exoplanet science outcomes of such a mission, we simulated the capabilities of a conceptual SMEX mission (EVE), which will obtain continuous NUV, optical, and NIR photometry for $\approx 20,000$ stars brighter than $T_\mathrm{mag} =16$ and younger than 50\,Myr over the course of 30 individual pointings, with observations spanning at least 30\,days per field. In this preliminary study, we show that if the majority of planets are gas-dwarfs born early in a gas-rich disk, then we expect to find $99\pm7$ planets over the primary mission survey. These transits will be primarily identified via joint analyses of the optical and NIR channels, and will be aided by flare modeling and mitigation from the NUV channel. If instead the planets form as water-worlds with high mean molecular weight envelopes then we expect very few planet $(5\pm2)$ detections, as all planets are expected to be born with radii $<3\,R_\oplus$. Significantly, the number of super-Neptunes that we detect is dependent on how depleted the gas disk is when these planets are formed, and a young planet detection mission has the potential to further refine our understanding of the gas accretion phase many planets must go through early in their formation process.

This sample of planets would provide the first set of significant constraints on the primordial planet frequency and distributions. We expect hydrogen/helium envelope gas-dwarfs to be Neptune-sized or larger, corresponding to the initial large radii expected for such young planets. We note the added science benefits of such a primordial planet population, including probing the relationship between inner disk structure and planet formation, constraining orbital evolution and migration models, and providing targets for atmospheric observations with JWST and Ariel. 

Alternatively, if water-worlds are predominant around M-dwarfs, then there is a possibility that hydrogen-ocean planets may be a standard outcome of planet formation. These hypothetical hydrogen ocean planets, such as K2-18b \citep{2003ApJ...596L.105K,2004Icar..169..499L}, have been speculated to host habitable conditions if a shallow ocean exists about a geologically active crust \citep{2015MNRAS.452.3752K,2018ApJ...864...75K,2021ApJ...918....1M}. 

\acknowledgements
We respectfully acknowledge the traditional custodians of all lands throughout Australia, and recognize their continued cultural and spiritual connection to the land, waterways, cosmos, and community. We pay our deepest respects to all Elders, ancestors and descendants of the Giabal, Jarowair, and Kambuwal nations, upon whose lands this research was conducted.
Part of this research was carried out at the Jet Propulsion Laboratory, California Institute of Technology, under a contract with the National Aeronautics and Space Administration (NASA).  
GZ thanks the support of the ARC DECRA program DE210101893 and ARC Future program FT230100517.
JGR gratefully acknowledges support from the Kavli Foundation. 
EJL was supported by NSF Research grant 2509275
AWM was supported by grants from the NSF CAREER program (AST-2143763) and NASA's exoplanet research program (XRP 80NSSC25K7148). 
MGB was supported by NSF Graduate Research Fellowship (DGE-2040435).
EG was supported by NASA award 80NSSC20K0957 (Exoplanets Research Program) and NASA’S Interdisciplinary Consortia for Astrobiology Research (NNH19ZDA001N-ICAR) under award number 19-ICAR19 2-0041. 

\appendix

\section{False positive predictions} \label{sec:falsepos}

Astrophysical false positives due to eclipsing binary systems are always identified alongside planets in wide field transit surveys. Without additional follow-up observations, these can often contaminate the planet statistics from a mission. We account for astrophysical signals from 
\begin{itemize}
    \item Unblended eclipsing binaries (EB) -- Often a F, G, K primary star with a short period transiting M-dwarf, with a radius ratio similar to that of hot Jupiters transiting equivalent hosts. 
    \item Hierarchical triples (HEB) -- Often a F, G, K primary star and a bound, fainter, eclipsing binary system (e.g. M and M pair). The eclipse depth is significantly diluted due to the flux ratio between the primary and the secondary binary, resulting in a shallow transit depth corresponding to a planetary transit.  
    \item Nearby background binary (NEB) -- A foreground target star with a background, unassociated eclipsing binary within the same pixel response function. The eclipse depth of the background binary is diluted by the foreground star, yielding a transit depth similar to a planetary transit.   
\end{itemize}

We follow \citet{sullivan15} to quantify the false positive statistics expected from EVE. For each target star that is $<50$ Myr old in our input catalog, 
\begin{enumerate}
    \item We compute the binary probability based on the multiplicity function from \citet{sullivan15} Table 3 as a function of stellar mass (in turn taken from \citealt{2004ASPC..318..166D, 2010ApJS..190....1R, 2013ARA&A..51..269D}).
    \item For target stars selected to be binaries, their companion mass ratio $q$ from the power law distribution $\frac{dN}{dq} \propto q^\gamma$, with $\gamma$ as a function of primary star mass and listed in \citet{sullivan15} Table 3. Similarly the orbital semi-major axis is drawn from a log normal distribution, with mean and standard deviations taken from \citet{sullivan15} Table 3. 
    \item Transiting systems are selected based on the transit probability $(R_1/a)$ of each system. The transit depth is simply computed to be $(R_2/R_1)^2$.
    \item Hierarchical triples are drawn based on the triple system multiplicity probability from \citet{2013ARA&A..51..269D}. These triples involve a primary star (star 1) orbited by close-in eclipsing binary system (stars 2 and 3). The period $P_\mathrm{1-23}$ and mass ratio $q_{1-23}$ of the lower mass eclipsing binary companion to the primary is drawn from the log normal and power law distributions prescribed above. The period of the eclipsing binary itself is drawn as per Eq 6 of \citet{sullivan15} $P_\mathrm{23} / P_\mathrm{1-23} = 0.2\times 10^{-2u}$, where $u$ is drawn uniformly from 0 to 1. The transit depth of hierarchical triples is computed to be the intrinsic transit depth $(R_3/R_2)^2$ diluted by the flux ratio between the primary and the secondary binary system $(F_2+F_3) / F_1$. 
    \item To determine the background binary statistics, we sample each survey field a \emph{Gaia} DR3 to determine the average stellar density and properties as a function of magnitude for this field. For each target star, we then randomly populate the area covered by the pixel response function with background stars fainter than the target based on the \emph{Gaia} derived stellar density. The binarity and eclipse depth of background stars are determined as per steps 1,2, and 3 via the multiplicity fraction stellar mass relationship. The eclipse depths are diluted based on the flux ratio between the target star and the background star. 
\end{enumerate}

The expected number of systems involving eclipsing binaries are tabulated in Table~\ref{tab:EBs}. We assume any system with transit depths of $<3\%$ can be confused with a transiting planet system. We also adopt an optimistic detection threshold of $\mathrm{SNR}>6$ for any such system. 

\begin{deluxetable*}{lrrr}
\tablewidth{0pc}
\tabletypesize{\scriptsize}
\tablecaption{
    Table of expected yield for eclipsing binaries 
    \label{tab:EBs}
}
\tablehead{ \\
    \multicolumn{1}{c}{Type}   &
    \multicolumn{1}{c}{Number detectable} &
    \multicolumn{1}{c}{Planet-like transits} &
    \multicolumn{1}{c}{NIR vs Optical color difference} \\
}
\startdata
    EB $<50$\,Myr & $28.2\pm3.5$ & $0.2\pm0.4$ & $0.0\pm0.0$\\
    HEB $<50$\,Myr & $42.5\pm4.2$ & $19.0\pm1.2$ & $9.3\pm1.1$\\
    NEB $<50$\,Myr & $43.0\pm3.4$ & $23.0\pm4.1$ & $2.5\pm1.5$\\
\enddata
\tablenotetext{a}{EB = eclipsing binary; HEB = hierarchical eclipsing binary; NEB = nearby eclipsing binary within the same PSF.}
\end{deluxetable*}

\subsection{Independent transit depths from optical and NIR channels}

A multi-band photometric survey will measure the transit depths of planet candidates in the optical and NIR. Depth differences may be expected due to specific blended astrophysical false positive scenarios. Figure~\ref{fig:opticalvsNIR} shows that the planet discoveries are expected to have similar transit signal-to-noise detections in the optical and NIR. In addition, all joint planet detections are expected to be detected in both optical and NIR at the $>5\sigma$ level, and all planets will have their radii constrained to similar precision across both channels. 

Hierarchical triples and nearby eclipsing binaries can exhibit color-dependence on the observed transit depth. In these scenarios, the flux ratio between the primary star and the eclipsing binary pair varies as a function of wavelength due to their differing effective temperatures. Planetary transits and two-body eclipsing binaries are largely gray in transit depths at the relevant precisions for discovery and detection. Multi-band transits are often obtained during the photometric follow-up campaign of transiting planet candidates to rule out such obvious false positive scenarios \citep[e.g.][]{2019AAS...23314005C,2021AJ....162..128W}. 

To determine the number of false positive detections that exhibit color-dependence in their transit depths detectable by the optical and infrared channels, we simulate the transits of all predicted HEB and NEB systems. For each system, we adopt fluxes from the MIST isochrones \citep{Choi2016} in $TESS$ band for the optical, and the summed flux in $J$ and $H$ bands for the NIR band. We find that 70\% of the false positives will exhibit significant depth difference between the optical and NIR bands at the $1\sigma$ level.



\bibliographystyle{aasjournal}
\bibliography{refs,references_master,ref_masterlist}

\end{document}